# Informe Técnico / Technical Report

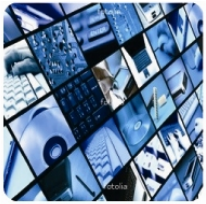 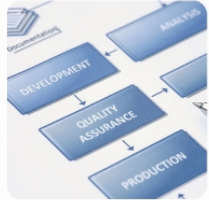 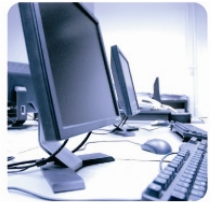 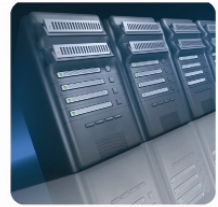 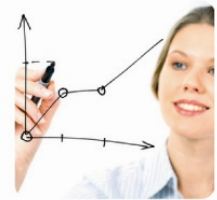

## Model-driven system development

### Experimental design and report of the pilot experiment

*Sergio España, Nelly Condori, Roel Wieringa, Arturo González, Óscar Pastor*

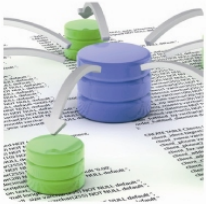 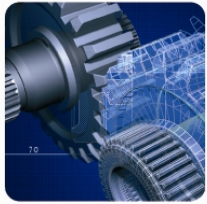 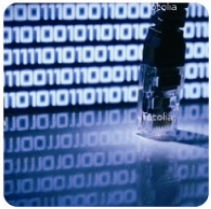 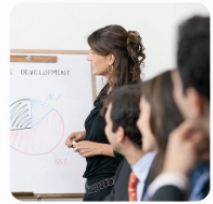 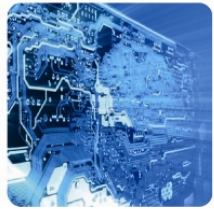

# MODEL-DRIVEN SYSTEM DEVELOPMENT

## EXPERIMENTAL DESIGN AND REPORT OF THE PILOT EXPERIMENT (HELD IN UNIVERSITY OF TWENTE IN 2009)

**Involved researchers**: Sergio España, Nelly Condori, Roel Wieringa, Arturo González, Óscar Pastor

**Corresponding author**: sergio.espana@pros.upv.es

Feel free to contact to request further information, to propose a collaboration (e.g. a replication package) or to provide feedback.

**Scope of this document**

This report describes de design of an experiment that intends to compare two variants of a model-driven system development method, so as to assess the impact of requirements engineering practice in the quality of the conceptual models. The conceptual modelling method being assessed is the OO-Method [Pastor and Molina 2007]. One of its variants includes Communication Analysis, a communication-oriented requirements engineering method [España, González et al. 2009] and a set of guidelines to derive conceptual models from requirements models [España, Ruiz et al. 2011; González, España et al. 2011]. The other variant is an ad-hoc, text-based requirements practice similar to the one that is applied in industrial projects by OO-Method practitioners. If you intend to cite this technical report to refer to the methods, please consider citing also/instead the following publications:

España, S., A. González and Ó. Pastor (2009). Communication Analysis: a requirements engineering method for information systems. 21st International Conference on Advanced Information Systems (CAiSE'09). Amsterdam, The Netherlands, Springer LNCS 5565: 530-545.

A. González, S. España, M. Ruiz, and Ó. Pastor, "Systematic derivation of class diagrams from communication-oriented business process models." In: 12th Working Conference on Business Process Modeling, Development, and Support (BPMDS'11), London, United Kingdom, 2011.

S. España, M. Ruiz, A. González, and O. Pastor, "Systematic derivation of state machines from communication-oriented business process models Derivation of the dynamic view of conceptual models in an industrial model-driven development method." In: Fifth IEEE International Conference on Research Challenges in Information Science (RCIS'11), Guadeloupe, France, 2011.

Pastor, O. and J. C. Molina (2007). Model-Driven Architecture in practice: a software production environment based on conceptual modeling. New York, Springer.

**TABLE OF CONTENTS**

# ACKNOWLEDGEMENTS

We, the researchers, are grateful to the following institutions and individuals; this research would not have been possible without them.

CARE for their strong support, especially to Juan Carlos Molina and José Miguel Barberá; for letting us use their technology for academic purposes, for creating the reference conceptual models, for collaborating in training the subjects. Elena Tejadillos for training the subjects in *OLIVANOVA*.

Bjorn Gerrit and Nazim Madhavji for their impressions and feedback on the experiment design.

Cristina Maestre for her graphic designs and her moral support.

Samuel Segarra for his definition of the master degree processes in which Problem 3 description was based.

Klaas Sikkel and Maya Daneva for their interest and support in making this experiment possible.

Suse Engbers for being in top of everything.

Ignacio Panach for carrying out the code-generation live demo.

Also, Sergio España wants to thank all the people that suported him in University of Twente, making his life easier and nicer there.

# 1. RESEARCH OBJECTIVES

The goal of the research, summarised according to the Goal/Question/Metric template [Basili and Rombach 1988], is to:

- **analyse** the resulting models of two model-based information systems analysis method variants; namely, the OO-Method (OOM) and the integration of Communication Analysis and the OO-Method (CA+OOM),
- **for the purpose of** carrying out a comparative evaluation
- **with respect to** performance of the subject and acceptance of the method;
- **from the viewpoint of** the information systems researcher
- **in the context of** bachelor students.

## 1.1. Purpose

The OO-Method [Pastor, Gómez et al. 2001] is an object-oriented system development method that has an industrial tool support in the *OLIVA**NOVA*** suite [CARE Technologies] (a technology that provides automatic code generation capabilities). The methodological core of the OO-Method is a conceptual model of the software system under construction.

As far as the researchers know[1], the current practice of industrial practitioners that use the OO-Method includes eliciting requirements via interviews and specifying them in an unstructured textual form (see Figure 1). As stated by a company manager, "there is one requirements engineering method per analyst"; he also stated that most analysts consider requirements specification a burden, since the documentation is only used as a contract with the client. In practice, the analysts create the requirements specification and the conceptual model concurrently. This information has been corroborated by at least one analyst.

Communication Analysis is a communication-oriented requirements engineering method for information systems development that offers a flow of activities and a requirements structure that aims to facilitate requirements elicitation and specification [España, González et al. 2009].

---

[1] The researchers have access to the staff of companies that develop the *OLIVA**NOVA*** suite and use it in real software development projects. In some meetings held in 2009, the researchers elicited information regarding the requirements engineering practice of the analysts. However, we acknowledge that a systematic ethnographical research would allow to discover the actual practice with more accuracy and reliability, but this has not been carried out.

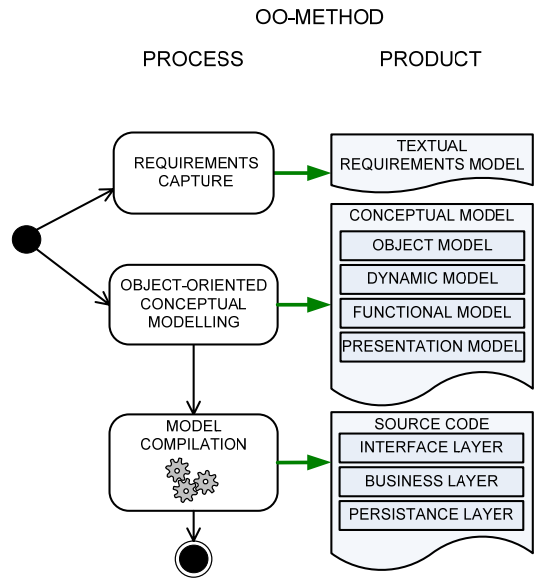

Figure 1. Overview of the OO-Method as used in current industrial practice

The OO-Method has been extended with a requirements phase based on Communication Analysis.

This research intends to assess to which extent the OO-Method benefits from its integration with Communication Analysis.

## 1.2. Object of study

To do this assessment, the researchers compare the outcomes of applying the two method variants (OO-Method and the integration of Communication Analysis and the OO-Method). Thus, the objects of study are the conceptual models (see Figure 2).

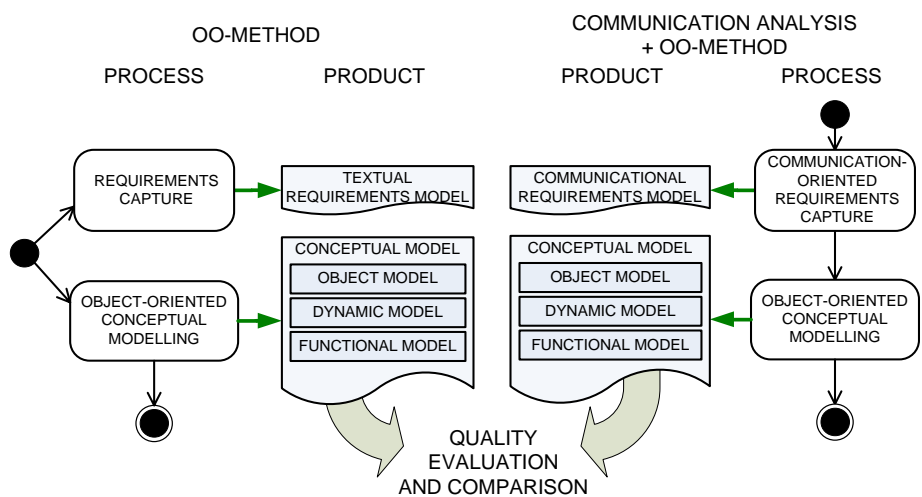

Figure 2. Overview of the comparison

The OO-Method conceptual model is composed of four views; namely the Object Model, the Dynamic Model, the Functional Model and the Presentation Model. The Presentation Model falls out of the scope of this research. The requirements models are neither compared.

## 1.3. Quality focus

As a general framework for method comparison, we adopt the Method Evaluation Model (MEM) [Moody 2003]. It defines a set of (aggregated) variables and causal relationships among them.

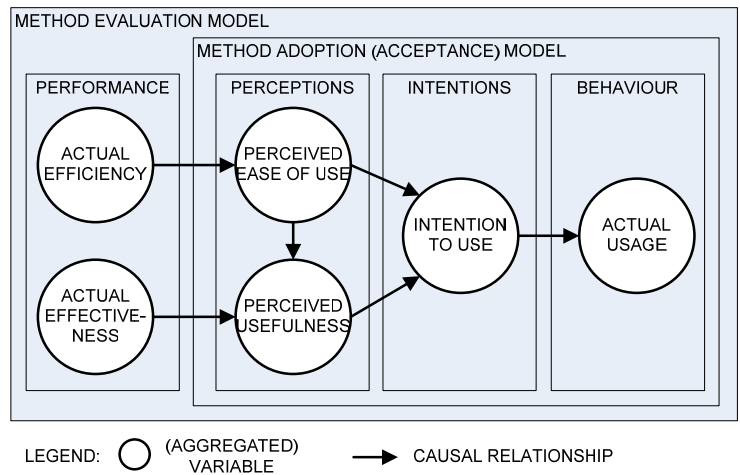

Figure 3.   Method Evaluation Model

The main effect studied in this research is the performance of the experimental subjects when applying the method variants to analyse information systems, and the acceptance the method variants have among the subjects. These effects are summarised next (for a more detailed description see Section 2.1.1).

*Performance* refers both to the effectiveness and the efficiency of the subjects in analysing creating the conceptual model that corresponds to the information system that an organisation needs.

*Efficiency* refers to the time it takes the subjects to create the conceptual model.

*Effectiveness* refers to the quality of the resulting conceptual models, when the subjects apply a method variant in order to solve a given problem. Quality is decomposed into three aspects: syntactic quality (syntactic correctness), semantic quality (feasible validity and feasible completeness) and pragmatic quality (feasible comprehension).

*Acceptance* refers both to the *perception* the subjects have with regard to the method variant (perceived ease of use and perceived usefulness) and their *intentions* towards them (intention to use)[2].

## 1.4. Perspective

The research is carried out from the point of view of the researchers. The researchers would like to know whether there is a significant difference between both method variants, in terms of the quality of the produced conceptual models. Based on previous experiences with both the OO-Method and Communication Analysis, the researchers expect that applying a communication-based requirements engineering method will improve some aspects of the conceptual model quality. From the point of view of the method designers, empirical evidences can be helpful for improving the methods.

---

[2] The *behaviour* of the subjects in the future (i.e. the actual usage of the method variants by the experimental subjects when they become industry practitioners) would require tracking them for several years and then assess whether they are actually using the method variants. We do not intend to do this.

## 1.5. Context

The expected subjects are 30 bachelor students of informatics and of business informatics (INF and BIT bachelor programs) at Twente University (Universiteit Twente, The Netherlands). They are studying to obtain their bachelor and they have finished their first year but may be in their second or third year.

The experiment is planned as part of a course that counts as Independent Project (*Vrij Project*), which means that it counts as 5 credits in the students curriculum. Thus, the selection of subjects from the population of all students is not the result of probability sampling but a convenience sampling; subjects freely choose to enrol in the course (and, therefore, to participate in the experiment).

## 1.6. Research questions

This research is designed as an experiment. The experiment addresses the following research questions:

- RQ1: Will the method variant influence the performance of experimental subjects?
- RQ2: Will both method variants have different acceptance among experimental subjects?
- RQ3: Will the modelling competence of experimental subjects influence their performance?
- RQ4: Will the modelling competence of experimental subjects influence their method variant acceptance?
- RQ5: Will there be a significant interaction between the modelling competence of experimental subjects and the method variant with regard to their performance?
- RQ6: Will there be a significant interaction between the modelling competence of experimental subjects and the method variant with regard to their acceptance of the method?
- RQ7: Will the MEM causal relationships appear in our experiment?

# 2. EXPERIMENT DESIGN

Each experimental subject solves three problems individually (Problems 1, 2 and 3 are the experimental objects). All subjects solve the three problems in the same order. Subjects are randomly[3] split into two groups (A and B). The method variant with which a subject solves each problem depends on the group they have been assigned (see Table 1). The result is a blocked subject-object study (according to [Wohlin, Runeson et al. 2000]).

|         | **Problem 1** | **Problem 2** | **Problem 3** |
|---------|---------------|---------------|---------------|
| **Group A** | OO-Method | OO-Method | Communication Analysis + OO-Method |
| **Group B** | OO-Method | Communication Analysis + OO-Method | Communication Analysis + OO-Method |

Table 1. Techniques applied to solve the problems.

The procedure to train the subjects in both method variants and have them solve the problems is depicted in Figure 4 (Section 3 further describes this procedure).

---

[3] The randomization procedure will be designed in a way that balance among groups is ensured.

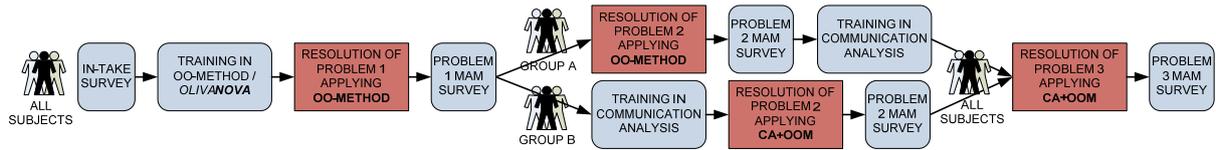

Figure 4. Overview of the experimental procedure

## 2.1. Variables of the experiment

We identify three types of variables [Juristo and Moreno 2001]: response variables (a.k.a. dependent variables), factors (a.k.a. independent variables) and parameters.

### 2.1.1. Response variables

Table 2 summarises the response variables considered in the experiment. Then, the variables are explained in detail.

The following mnemonic is used to refer to the instrumentation that supports the measurement: RAT stands for rating on a Likert scale, QUE stands for list of questions, and STA stands for list of statements.

Note that # is an abbreviation for number.

| Type and subtype | | | Variable | Metric |
|---|---|---|---|---|
| Performance | Efficiency | | Time to finish task | *Time* = # of hours spent on the creation of the conceptual model, reported by the user |
| | Effectiveness | Syntactic quality | Syntactic correctness | *Correctness_RAT* = Rating of the evaluator on a 7-point Likert scale |
| | | Semantic quality | Feasible validity | *Validity_RAT* = Rating of the evaluator on a 7-point Likert scale |
| | | | | *Validity_QUE* = 1 − (# of questions answered wrongly / # questions) |
| | | | | *Validity_STA* = 1 − (# of statements that are incorrectly specified in the model / # statements) |
| | | Pragmatic quality | Feasible completeness | *Completeness_RAT* = Rating of the evaluator on a 7-point Likert scale |
| | | | | *Completeness_QUE* = # of questions answered correctly / # questions |
| | | | | *Completeness_STA* = # of statements that are correctly specified in the model / # statements |

| | | Feasible comprehension | *Comprehension_RAT* = Rating of the evaluator on a 7-point Likert scale |
| --- | --- | --- | --- |
| | | | *Comprehension_QUE* = mean of Likert scale values (only considering correct and incorrect questions, not those questions for which the model does not provide an answer) |
| | | | *Comprehension_STA* = mean of Likert scale values (only considering correctly and incorrectly specified statements, not those statements not specified in the model) |
| Acceptance | Perception-based | Perceived ease of use | *PEOU_RAT* = mean of the Likert scale values of questionnaire items related to Perceived ease of use |
| | | Perceived usefulness | *PU_RAT* = mean of the Likert scale values of questionnaire items related to Perceived usefulness |
| | Intention-based | Intention to use | *ITU_RAT* = mean of the Likert scale values of questionnaire items related to Intention to use |

Table 2. Response variables considered in the experiment

#### 2.1.1.1. Performance variables

With regard to performance, we differentiate efficiency and effectiveness.

##### 2.1.1.1.1. *Variables related to efficiency*

We consider the time it takes the subject to create the conceptual model. The following means of measuring efficiency is considered:

- *Time* = # of hours spent on the creation of the conceptual model, reported by the user

##### 2.1.1.1.2. *Variables related to effectiveness*

Lindland, Sindre and Sølvberg [1994] present a framework of conceptual model quality that decomposes quality into three quality types: syntactic quality, semantic quality and pragmatic quality.

Syntactic quality is the extent to which the model does not contain flaws; the authors define one goal for syntactic quality: syntactic correctness is the degree to which the statements of the model adhere to the modelling grammar. The more the deviations from the rules (i.e. the more flaws), the less the syntactic quality.

Semantic quality is the degree to which the model correctly represents the problem domain; it has two goals: feasible validity is the extent to which the model does not include incorrect statements about the domain, and feasible completeness is the extent to which the relevant statements about the domain are included in the model.

Pragmatic quality is related to the audience of a model; it has one goal: feasible comprehension is the extent to which the model is understood by its intended readers.

There are at least three different means of measuring these variables.

    NOTES:

Moody, Sindre et al. [2002] [2003] did not decompose semantic quality into feasible completeness and feasible validity when carrying out their experiments, but we will.

Also, we will be measuring each quality variable using three different metrics (with their specific way of measuring). Using more than one metric allows us to analyse their correlation, but it raises the doubt about which metric serves as reference (i.e. which metric measures quality more accurately). For instance, if the results of one metric contradict the results of another metric and, moreover, if one of the results refutes our expected outcomes, then which results would we consider to be more reliable? (i.e. which results reflect reality more accurately?).

**Ratings on a 7-point Likert scale value** (RAT)

Following the approach by Moody, Sindre et al. [2002] [2003], in order to measure syntactic quality, semantic quality and pragmatic quality, the reviewer rates the model using a 7-point Likert scale.

- Syntactic quality
    - Syntactic correctness
        - *Correctness_RAT* = Rating of the evaluator on a 7-point Likert scale
- Semantic quality
    - Feasible validity
        - *Validity_RAT* = Rating of the evaluator on a 7-point Likert scale
    - Feasible completeness
        - *Completeness_RAT* = Rating of the evaluator on a 7-point Likert scale
- Pragmatic quality
    - Feasible comprehension
        - *Comprehension_RAT* = Rating of the evaluator on a 7-point Likert scale

**Problem-related questions** (*QUE*)

A meta-reviewer provides the reviewer with a list of questions related to the problem. For each question, the reviewer looks for the answer in the model. If they find the answer, they write it down, if not, they indicate so. Furthermore, the reviewer rates on a 7-point Likert scale how easy it was to understand the model while seeking the answer in it. Then, the meta-reviewer checks whether the answers provided by the reviewer coincide with the values that the domain dictates: if so, the question is considered to be answered correctly by the model; if not, the answer is considered to be answered wrongly by the model. Figure 5 shows a trivial example.

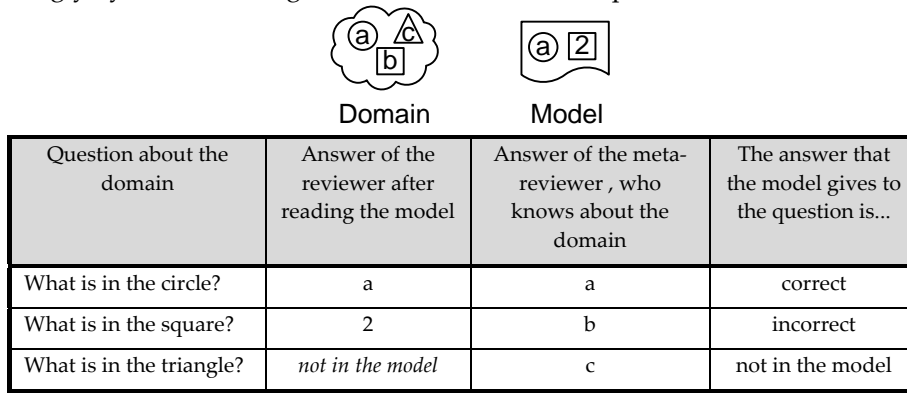

| Question about the domain | Answer of the reviewer after reading the model | Answer of the meta-reviewer, who knows about the domain | The answer that the model gives to the question is... |
|---|---|---|---|
| What is in the circle? | a | a | correct |
| What is in the square? | 2 | b | incorrect |
| What is in the triangle? | *not in the model* | c | not in the model |

Figure 5. Example of problem-related questions

Then it can be considered that feasible validity is inversely proportional to the number of incorrect answers, feasible completeness is directly proportional to the number of correct answers, and comprehension is directly proportional to the ease of understanding that the reviewer rated for the questions.

- Semantic quality
    - Feasible validity

- *Validity_QUE* = 1 − (# of questions answered wrongly / # questions)
    - Feasible completeness
        - *Completeness_QUE* = # of questions answered correctly / # questions
- Pragmatic quality
    - Feasible comprehension
        - *Comprehension_QUE* = mean of Likert scale values (only considering correct and incorrect questions, not those questions for which the model does not provide an answer)

**List of statements** (*STA*)

A meta-reviewer provides the reviewer with a list statements about the problem. The reviewer goes through the list and checks whether each statement is correctly specified in the model, incorrectly specified, or not specified at all. This procedure is similar to problem-related questions except that the instrumentation does not express questions but statements; that is, the answer is given a priori. Furthermore, the reviewer rates on a 7-point Likert scale how easy it was to understand the model while seeking the specification of the statement in it.

In the following, a simple example is explained, with the intention to clarify the list-based approach. Figure 6 shows a trivial example.

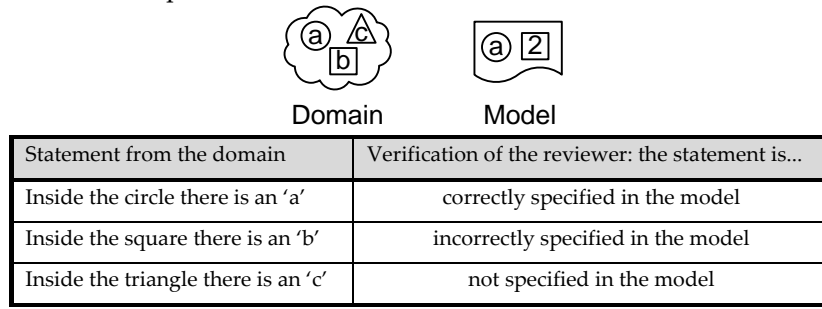

| Statement from the domain | Verification of the reviewer: the statement is... |
|---|---|
| Inside the circle there is an 'a' | correctly specified in the model |
| Inside the square there is an 'b' | incorrectly specified in the model |
| Inside the triangle there is an 'c' | not specified in the model |

Figure 6.   Example of a list of statements

Then it can be considered that feasible validity is inversely proportional to the number of incorrectly specified statements, feasible completeness is directly proportional to the number of correctly specified statements, and comprehension is directly proportional to the ease of understanding that the reviewer rated for the statements.

- Semantic quality
    - Feasible validity
        - *Validity_STA* = 1 − (# of statements that are incorrectly specified in the model / # statements)
    - Feasible completeness
        - *Completeness_STA* = # of statements that are correctly specified in the model / # statements
- Pragmatic quality
    - Feasible comprehension
        - *Comprehension_STA* = mean of Likert scale values (only considering correctly and incorrectly specified statements, not those statements not specified in the model)

**A note about the distinction between static and dynamic aspects of the problem domain.**

Validity, completeness and comprehension can be measured for both the static and the dynamic aspects of the domain. For instance:

**Ratings on a 7-point Likert scale value** (RAT)

- *Validity_static_RAT* = Rating of the evaluator on a 7-point Likert scale
- *Validity_dynamic_RAT* = Rating of the evaluator on a 7-point Likert scale
- *Completeness_static_RAT* = Rating of the evaluator on a 7-point Likert scale
- *Completeness_dynamic_RAT* = Rating of the evaluator on a 7-point Likert scale
- *Comprehension_static_RAT* = Rating of the evaluator on a 7-point Likert scale
- *Comprehension_dynamic_RAT* = Rating of the evaluator on a 7-point Likert scale

**Problem-related questions** (QUE)
- *Validity_static_QUE* = 1 − (# of questions addressing static aspects of the domain that are answered wrongly / # questions addressing static aspects)
- *Validity_dynamic_QUE* = 1 − (# of questions addressing dynamic aspects of the domain that are answered wrongly / # questions addressing dynamic aspects)
- *Completeness_static_QUE* = # of questions addressing static aspects of the domain that are answered correctly / # questions addressing static aspects of the domain
- *Completeness_dynamic_QUE* = # of questions addressing dynamic aspects of the domain that are answered correctly / # questions addressing dynamic aspects of the domain
- *Comprehension_static_QUE* = mean of Likert scale values for questions addressing static aspects of the domain (only considering correct and incorrect questions, not those questions for which the model does not provide an answer)
- *Comprehension_dynamic_QUE* = mean of Likert scale values for questions addressing dynamic aspects of the domain (only considering correct and incorrect questions, not those questions for which the model does not provide an answer)

**List of statements** (STA)
- *Validity_static_STA* = 1 − (# of statements related to static aspects of the domain that are incorrectly reflected in the model / # statements related to static aspects of the domain)
- *Validity_dynamic_STA* = 1 − (# of statements related to dynamic aspects of the domain that are incorrectly reflected in the model / # statements related to dynamic aspects of the domain)
- *Completeness_static_STA* = # of statements related to static aspects of the domain that are reflected in the model / # statements related to static aspects of the domain
- *Completeness_dynamic_STA* = # of statements related to dynamic aspects of the domain that are reflected in the model / # statements related to dynamic aspects of the domain
- *Comprehension_static_STA* = mean of Likert scale values statements related to static aspects of the domain (only considering correctly and incorrectly specified statements, not those statements not specified in the model)
- *Comprehension_dynamic_STA* = mean of Likert scale values for statements related to dynamic aspects of the domain (only considering correctly and incorrectly specified statements, not those statements not specified in the model)

However, in this experiment we do not plan to make this distinction.

### 2.1.1.2. Acceptance variables

We adopt the Method Evaluation Model (MEM) and we use it as in in [Moody, Sindre et al. 2002] and [Moody, Sindre et al. 2003]. The students answer a post-task survey that contains questions related to each of the following variables. Their answer to these questions is a rating on a 5-point Likert scale. MEM distinguishes between perception-based variables and intention-based variables.

#### 2.1.1.2.1. Perception-based variables

- Perceived ease of use: `PEOU_RAT` = mean of the Likert scale values of questionnaire items related to PEOU
- Perceived usefulness: `PU_RAT` = mean of the Likert scale values of questionnaire items related to PU

#### 2.1.1.2.2. Intention-based variables

- Intention to use: `ITU_RAT` = mean of the Likert scale values of questionnaire items related to ITU

### 2.1.2. Factors

The following factors are identified:

- Method variant. The variant of the systems analysis method applyed by the subject to engineer requirements and develop a conceptual model is the main factor in which we are interested in this research. Two treatments are considered: (a) the method that integrates Communication Analysis and OO-Method (CA+OOM) and (b) the OO-Method (OOM). The variable is named `Method_variant`.
- Modelling competence. The competence of the experimental subject in applying the OO-Method. This factor is a continuum ranging from 0 to 10 and it is valued by means of a knowledge level assessment (a test) and a modelling task. The variable is named `Competence_OOM`. A value from 5 to 7,5 means that the subject has an average competence level; a value greater that 7,5 means that the subject has a high competence level.

The following factors (a.k.a. independent variables) are identified:

### 2.1.3. Parameters

Variables that we do not want to influence the experimental results have been fixed:

- Problem domain: information systems analysis.
- Problem-size complexity: three problem statements of a similar size are chosen.
- Background of the experimental subjects: students with a similar background are chosen; however, their homogeneity will be verified by means of an intake knowledge level assessment.

## 2.2. Formulation of hypotheses

Research questions can be now specified as precise hypotheses.

Note that, some definitions actually represent several hypotheses, since _X stands for either of the metrics related to the variables. For instance, `Correctness_X` represents Correctness`_RAT`, Correctness`_QUE` and Correctness`_STA`.

**RQ1: Will the method variant influence the performance of experimental subjects?**

**H1.1**: There is a significant difference between the values of *Time* for both method variants.

We expect it takes the subjects less time to build the conceptual model when applying OOM than when applying CA+OOM.

**H1.2**: There is a significant difference between the values of *Correctness_X* for both method variants.

We expect that conceptual models built by applying CA+OOM are significantly more correct than conceptual models built by applying OOM.

**H1.3**: There is a significant difference between the values of *Validity_X* for both method variants.

We expect that conceptual models built by applying CA+OOM are significantly more valid than conceptual models built by applying OOM.

**H1.4**: There is a significant difference between the values of *Completeness_X* for both method variants.

We expect that conceptual models built by applying CA+OOM are significantly more complete than conceptual models built by applying OOM.

**H1.5**: There is a significant difference between the values of *Comprehension_X* for both method variants.

We expect that conceptual models built by applying CA+OOM are significantly more comprehensible than conceptual models built by applying OOM.

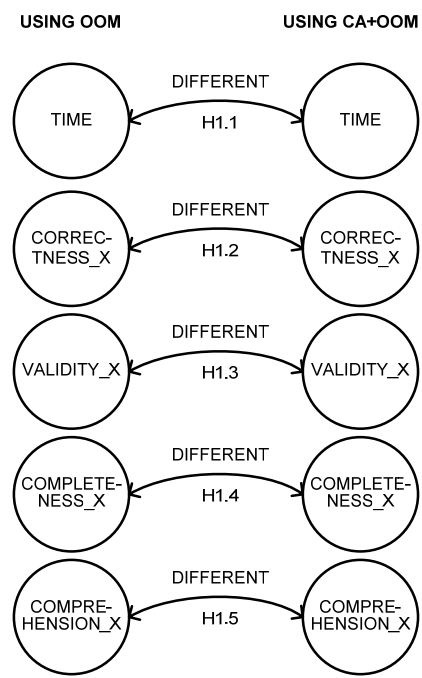

Figure 7. Graphical representation of hypotheses related to RQ1

**RQ2: Will both method variants have different acceptance among experimental subjects?**

**H2.1**: There is a significant difference between the values of *PEOU_RAT* for both method variants.

We expect that subjects significantly perceive OOM as more easy to use than CA+OOM.

**H2.2**: There is a significant difference between the values of *PU_RAT* for both method variants.

We expect that subjects significantly perceive CA+OOM as more useful than OOM.

**H2.3**: There is a significant difference between the values of *ITU_RAT* for both method variants.

We do not expect a significant difference between the intentions of the subjects with regards to using any of the method variants in the future.

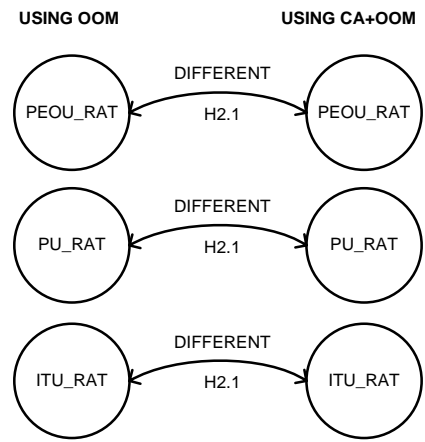

Figure 8. Graphical representation of hypotheses related to RQ2

**RQ3: Will the modelling competence of experimental subjects influence their performance?**

**H3.1**: *Competence_OOM* correlates with *Time*.

We expect that the more modelling competence in the OO-Method a subject has, the less time it takes them to build the conceptual model.

**H3.2**: *Competence_OOM* correlates with *Correctness_X*.

We expect that the more modelling competence in the OO-Method a subject has, the more correct their model is.

**H3.3**: *Competence_OOM* correlates with *Validity_X*.

We expect that the more modelling competence in the OO-Method a subject has, the more valid their model is.

**H3.4**: *Competence_OOM* correlates with *Completeness_X*.

We expect that the more modelling competence in the OO-Method a subject has, the more complete their model is.

**H3.5**: *Competence_OOM* correlates with *Comprehension_X*.

We expect that the more modelling competence in the OO-Method a subject has, the more comprehensive their model is.

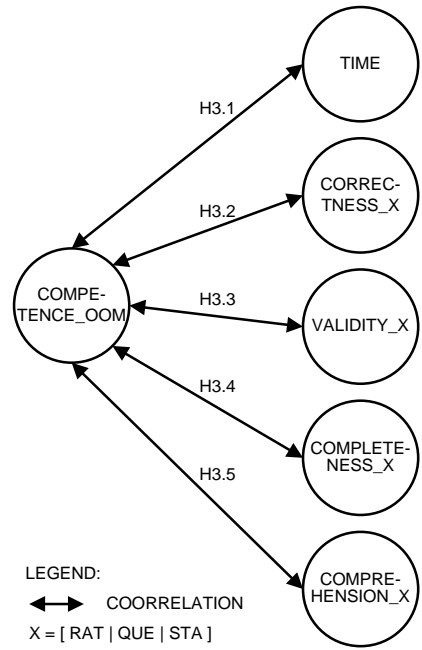

Figure 9. Graphical representation of hypotheses related to RQ3

**RQ4: Will the modelling competence of experimental subjects influence their method variant acceptance?**

**H4.1**: *Competence_OOM* correlates with *PEOU_RAT*.

We expect that the more modelling competence in the OO-Method a subject has, the more they perceive the method variant to be easy to use.

**H4.2**: *Competence_OOM* correlates with *PU_RAT*.

We expect that the more modelling competence in the OO-Method a subject has, the more they perceive the method variant to be useful.

**H4.3**: *Competence_OOM* correlates with *ITU_RAT*.

We expect that the more modelling competence in the OO-Method a subject has, the more they intend to use the method variant in the future.

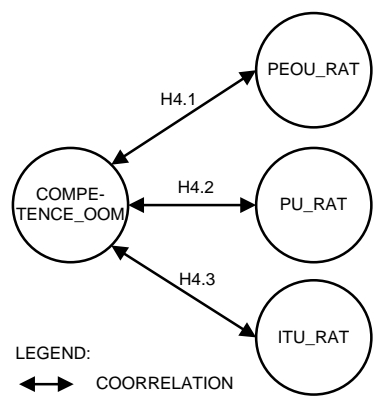

Figure 10. Graphical representation of hypotheses related to RQ4

**RQ5: Will there be a significant interaction between the modelling competence of experimental subjects and the method variant with regard to their performance?**

**H5.1**: The difference between the values of *Time* for experimental subjects with high modelling competence that used *Method_variant* OOM and for experimental subjects with high modelling competence that used *Method_variant* CA+OOM is significantly different to the difference between the values of *Time* for experimental subjects with low modelling competence that used *Method_variant* = OOM and for experimental subjects with low modelling competence that used *Method_variant* = CA+OOM.

We do not expect a significant interaction between the OO-Method modelling competence and the method variant with respect to the time it takes to build the conceptual model.

**H5.2**: The difference between the values of *Correctness_X* for experimental subjects with high modelling competence that used *Method_variant* OOM and for experimental subjects with high modelling competence that used *Method_variant* CA+OOM is significantly different to the difference between the values of *Correctness_X* for experimental subjects with low modelling competence that used *Method_variant* OOM and for experimental subjects with low modelling competence that used *Method_variant* CA+OOM.

We expect that the effect of CA is more pronounced for subjects with bad modelling competence than for subjects with good modelling competence.

**H5.3**: Idem for *Validity_X*.

We expect that the effect of CA is more pronounced for subjects with bad modelling competence than for subjects with good modelling competence.

**H5.4**: Idem for *Completeness_X*.

We expect that the effect of CA is more pronounced for subjects with bad modelling competence than for subjects with good modelling competence.

**H5.5**: Idem for *Comprehension_X*.

We expect that the effect of CA is more pronounced for subjects with bad modelling competence than for subjects with good modelling competence.

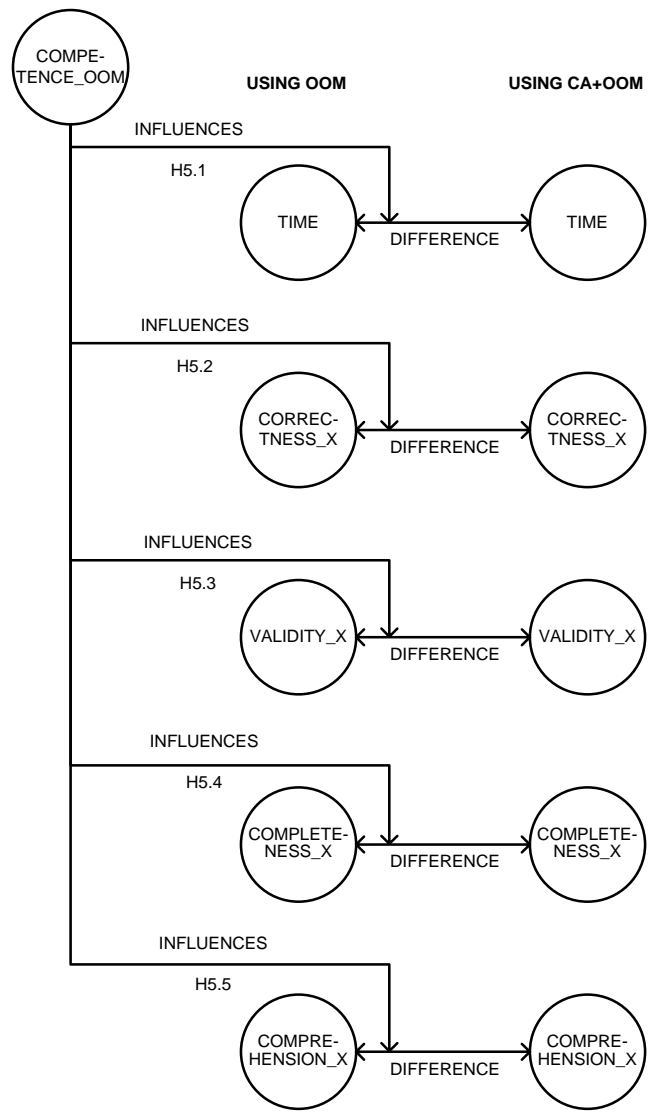

Figure 11. Graphical representation of hypotheses related to RQ5

**RQ6: Will there be a significant interaction between the modelling competence of experimental subjects and the method variant with regard to their acceptance of the method?**

**H6.1**: The difference between the values of *PEOU_RAT* for experimental subjects with high modelling competence that used *Method_variant* OOM and for experimental subjects with high modelling competence that used *Method_variant* CA+OOM is significantly different to the difference between the values of *PEOU_RAT* for experimental subjects with low modelling competence that used *Method_variant* OOM and for experimental subjects with low modelling competence that used *Method_variant* CA+OOM.

We do not expect a significant interaction.

**H6.2**: Idem for *PU_RAT*.

We do not expect a significant interaction.

**H6.3**: Idem for *ITU_RAT*.

We do not expect a significant interaction.

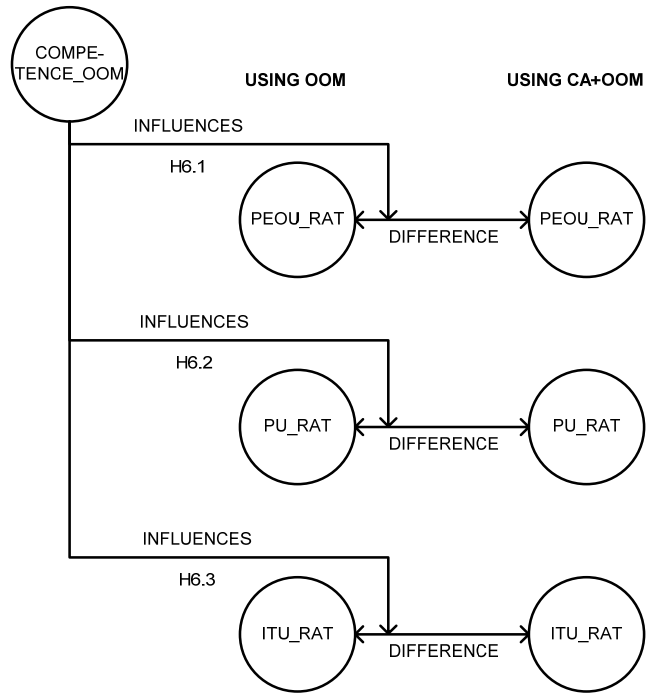

Figure 12. Graphical representation of hypotheses related to RQ6

**RQ7: Will the MEM causal relationships appear in our experiment?**

**H7.1**: *TIME* correlates with *PEOU_RAT*.

**H7.2**: *Correctness_X* correlates with *PU_RAT*.

**H7.3**: *Validity_X* correlates with *PU_RAT*.

**H7.4**: *Completeness_X* correlates with *PU_RAT*.

**H7.5**: *Comprehension_X* correlates with *PU_RAT*.

**H7.6**: *PEOU_RAT* correlates with *PU_RAT*.

**H7.7**: *PEOU_RAT* correlates with *ITU_RAT*.

**H7.8**: *PU_RAT* correlates with *ITU_RAT*.

We expect all this correlations to appear, although some of them could be non significant.

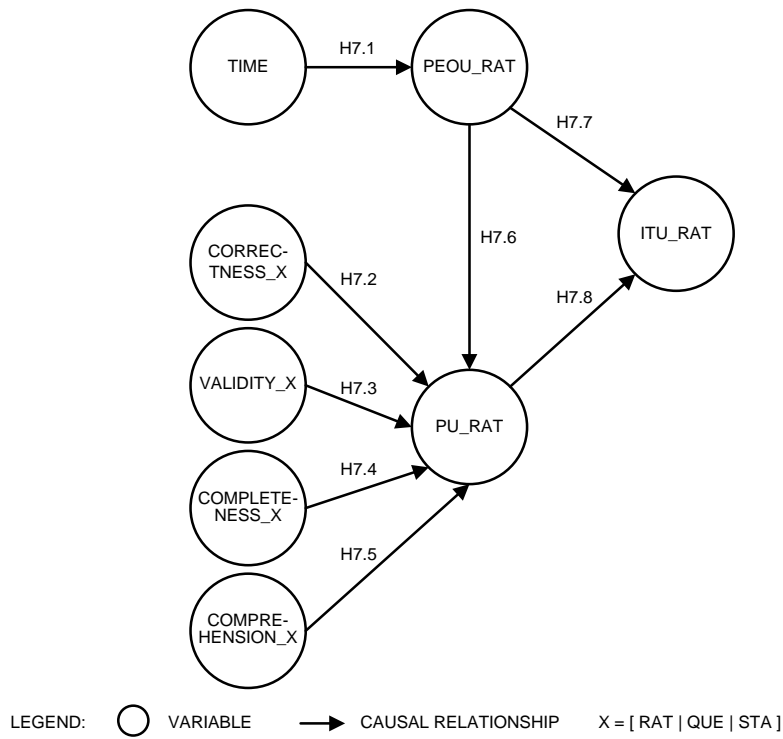

Figure 13. Graphical representation of hypotheses related to RQ7

Note that the MEM has been extended in order to adapt it to our experiment (see Figure 14). We intend to test some additional correlations (see greyed variables and arrows).

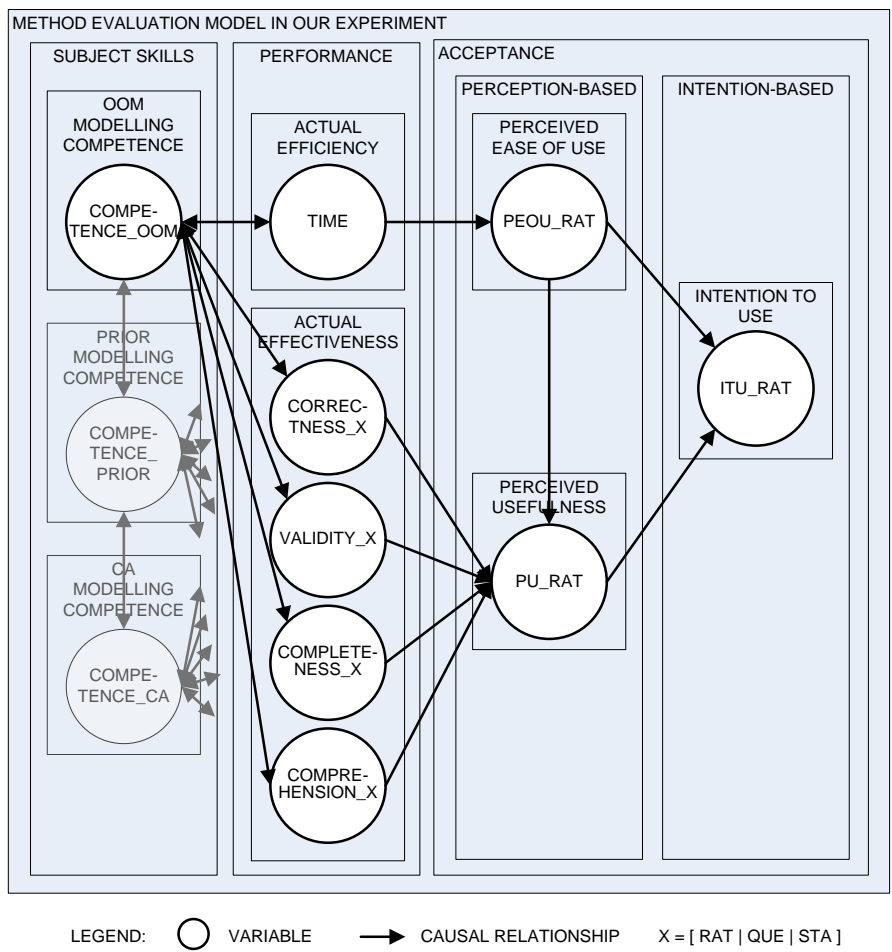

Figure 14. Method Evaluation Model in our experiment

# 3. EXPERIMENTAL PROCEDURE

The planning of the experiment is depicted in Figure 15 (simplified view) and in Figure 16 (detailed view). Table 3 shows the expected dedication of the experimental subjects; the big amount of hours that the subjects devote to the experiment makes this a mid-term experiment (at least compared to the usual requirements engineering experiments) and the fact that the experimental tasks are part of the students homework implies that we loose some control over their way of working (and, thus, over the treatment), but this seems unavoidable in this kind of setting.

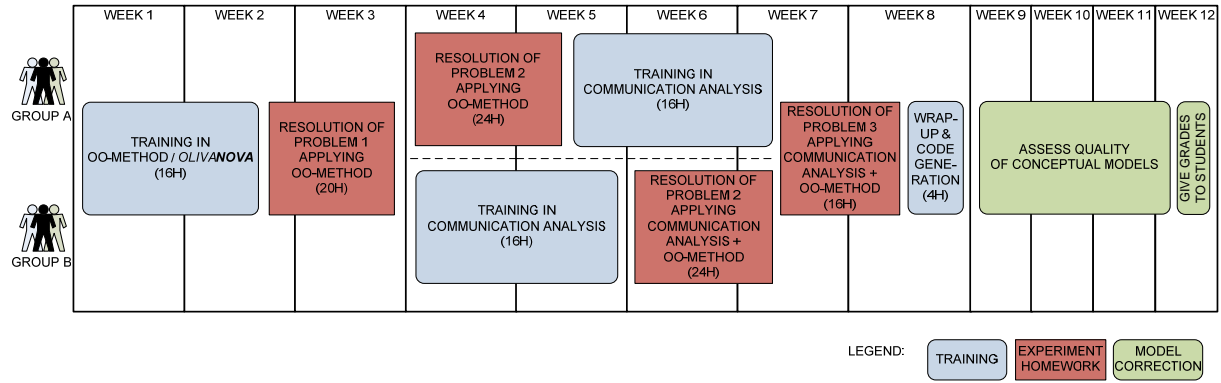

Figure 15. Planning of the experiment (simplified)

| | Monday | Tuesday | Wednesday | Thursday | Friday | Saturday | Sunday |
|---|---|---|---|---|---|---|---|
| **31/08/2009 - 25/10/2009** | | | | | | | |
| **WEEK 1** | August 31 | September 1 | 2 | 3 | 4 | 5 | 6 |
| | | | KICK-OFF (1H) AB [TEST] / OOM INTRO (1H) AB / OOM OBJECT (2H) AB | OOM OBJECT (4H) AB / OOM OBJECT (2H) AB / OOM DYNAMIC (2H) AB | OOM DYNAMIC (4H) AB | | H of class / homework  A:  8  8  B:  8  8 |
| **WEEK 2** | 7 | 8 | 9 | 10 | 11 | 12 | 13 |
| | OOM DYNAM. (2H) AB / OOM FUNCT. (2H) AB | OOM FUNCT. (4H) AB | OOM FUNCT. (2H) AB / OOM SUM-UP (1H) AB [TEST] | PROBLEM 1 OOM (4H) AB | | | H of class / homework  A:  8  8  B:  8  8 |
| **WEEK 3** | 14 | 15 | 16 | 17 | 18 | 19 | 20 |
| | PROBLEM 1 OOM (16H) AB [SURVEY] | | | OOM DOUBTS (2H) AB | | | H of class / homework  A:  2  16  B:  2  16 |
| **WEEK 4** | 21 | 22 | 23 | 24 | 25 | 26 | 27 |
| Gr.A | PROBLEM 2 (16H) A | | | | | | H of class / homework  A:  -  16  B:  8  8 |
| Gr.B | CA INTRO (2H) B / CA PROCESS (2H) B | CA PROCESS (4H) B | CA PROCESS (2H) B / CA COMMUN. (2H) B | CA COMMUN. (4H) B | | | |
| **WEEK 5** | 28 | 29 | 30 | October 1 | 2 | 3 | 4 |
| Gr.A | PROBLEM 2 (8H) A [SURVEY] | | CA INTRO (2H) A / CA PROCESS (2H) A | CA PROCESS (4H) A | | | H of class / homework  A:  4  12  B:  8  8 |
| Gr.B | CA COMMUN (2H) B / CA INTEG OOM (2H) B | CA INTEG OOM (8H) B | | CA INTEG OOM (4H) B [TEST] | | | |
| **WEEK 6** | 5 | 6 | 7 | 8 | 9 | 10 | 11 |
| Gr.A | CA PROCESS (2H) A / CA COMMUN. (2H) A | CA COMMUN. (4H) A | CA COMMUN (2H) A / CA INTEG OOM (2H) A | CA INTEG OOM (8H) A | | | H of class / homework  A:  8  12  B:  -  16 |
| Gr.B | PROBLEM 2 (16H) B | | | | | | |
| **WEEK 7** | 12 | 13 | 14 | 15 | 16 | 17 | 18 |
| Gr.A | CA INTEG OOM (4H) A [TEST] | | CA DOUBTS (2H) A?B | PROBLEM 3 (8H) AB | | | H of class / homework  A:  6  8  B:  2  16 |
| Gr.B | PROBLEM 2 (8H) B [SURVEY] | | | | | | |
| **WEEK 8** | 19 | 20 | 21 | 22 | 23 | 24 | 25 |
| | PROBLEM 3 (8H) AB [SURVEY] | | CODE GENER. (2H) AB / WRAP-UP (2H) AB [SURVEY] | | | | H of class / homework  A:  4  8  B:  4  8 |

LEGEND: THEORY CLASS | PRACTICE CLASS | NORMAL HOMEWORK | EXPERIMENT HOMEWORK

Figure 16. Planning for the experiment (detailed)

| Group | A | | B | |
|---|---|---|---|---|
| Week | Class | Homework | Class | Homework |
| 1 | 8 | 8 | 8 | 8 |
| 2 | 8 | 8 | 8 | 8 |
| 3 | 2 | 16 | 2 | 16 |
| 4 | 0 | 16 | 8 | 8 |
| 5 | 4 | 12 | 8 | 8 |
| 6 | 8 | 12 | 0 | 16 |
| 7 | 6 | 8 | 2 | 16 |
| 8 | 4 | 8 | 4 | 8 |
| Total per type | 40 | 88 | 40 | 88 |
| Total | 128 | | 128 | |

Table 3. Expected dedication of students to the course/experiment (in hours)

## 3.1. Kick-off

The subjects are informed about the nature of the research, they are given an overview of the experiment, they are explained their expected dedication and workload.

In this meeting, the subjects fill out an **in-take survey** (see Section 4.1.1) that aims to assess their modelling competence (we differentiate static and dynamic aspects) before they attend the training seminars[4].

Also, the *OLIVANOVA* Modeler is distributed to the subjects along with instructions for installing the software.

## 3.2. Training on OO-Method / OLIVANOVA

The subjects are trained on the OO-Method, during a course composed of several seminars[5]. Table 4 presents the outline of the course (for a description of the **training material**, see Section 4.2.1).

| Seminar # | Content |
|---|---|
| 1 | Introduction to *OLIVANOVA* The Programming Machine |
| | Classes & Relationships |
| | Attributes |
| | Object Model exercise |
| 2 | Services |

---

[4] Those subjects with a higher level of knowledge should be closely monitored, since they may be regarded as 'experts' (they could perform particularly well and offer valuable qualitative data).

[5] The reinforcement pedagogical pattern [Berenbach and Konrad 2008] is used for running the seminars: each concept is first defined, then the trainer offers one or more illustrative examples of the concept, then one or more class exercises are proposed (students volunteer to answer questions and solve toy-size problems, and the trainer fosters debate). After a set of related concepts has been introduced, a team exercise is proposed (class is split into teams, the problem takes from 20 to 40 minutes to be solved, a spokesman from each team presents the team solution, and the whole class debates the proposed solutions). The aim of the course is to achieve at least the first five categories of Bloom's taxonomy of educational objectives [Bloom, Engelhart et al. 1956] (namely; knowledge, comprehension, application, analysis and synthesis).

| | Functional Model |
| | Transactions and operations |
| | Functional Model exercise |
| 3 | Agents |
| | Preconditions and Integrity Constraints |
| | State Transition Diagram |
| | Dynamic Model exercise |
| 4 | Validation |
| | *OLIVANOVA* Modeler Options and additional features |

Table 4. Outline of the OO-Method training

During the course, a representative example of an Information System conceptual modelling is developed. The subjects can use both the course material and the representative example as references during later experimental tasks.

After the OO-Method training, the knowledge level of the subjects is assessed (the aim is to assess the degree of achievement of Bloom's educational objectives). The OO-Method modelling competence assessment consists of a test multiple-choice questions (the *OLIVANOVA* **knowledge test**, see Section 4.2.2) and a small size conceptual modelling exercise (the *OLIVANOVA* **modelling exercise**, see Section 4.2.3). Both the test and the problem statement are part of the training material developed by CARE Technologies, although they have been slightly adapted for this experiment[6]. Then, for correcting the modelling exercise, a *OLIVANOVA* **modelling exercise correction template** has been designed (see Section 4.2.4).

This assessment allows giving value to the variable *Competence_OOM*. Also, if the test results indicate that one or several methodological concepts (or their related modelling primitives) are considerably misunderstood, additional clarification may be offered (i.e. extended course material may be handed out to each and every subject).

After this course, Problem 1 is delivered, along with conceptual modelling task instructions.

## 3.3. Resolution of Problems 1,2,3

Problem 1, Problem 2 and Problem 3 are the three experimental tasks.

The subjects are given the **task instructions** (see Section 4.3.2), which is a guide describing the experimental task and the different protocols to be followed: how to elicit requirements, how and when to deliver a snapshot[7], the schedule of follow-up meetings, the expected outcomes of the experimental task, the tools they are expected to use (e.g. a diagramming tool in order to enhance the comprehensibility of models), etc.

---

[6] The original test is longer, since it includes questions about the Presentation Model; also, many questions have been rephrased and it has been supported by means of a web-based form to facilitate the gathering and correction of responses.

The problem statement involves some attributes more, and an additional agent of the system, but it has been simplified to fit the allotted time.

[7] A snapshot is a copy of the models (requirements models and/or conceptual models) being produced by a subject as they are at a given moment in time. Snapshots allow the experimenters to monitor the evolution of the subject's work and are also intended to motivate them to work regularly during the alloted time and prevent last-minute

Subjects are first given a **short problem description** (see Section 4.3.3). This statement only gives an overview of the organisation and its work practice; it does not include each and every requirement of the information system. The subjects need to elicit requirements in order to discover the problem. In order to offer an unbiased and replicable source of requirements, the subjects elicit requirements via a requirements **helpdesk system** (see Section 4.3.1).

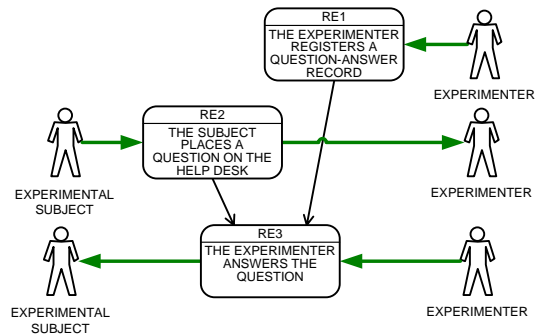

Figure 17.  Communicative event diagram of the help-desk-based elicitation

As shown in Figure 17, whenever a subject needs to know something about the problem at hand, they can place questions via the helpdesk. The experimenter looks up each question in a question-answer catalogue and sends the answers to the subject (again, via the helpdesk). In case the question is not found in the **question-answer catalogue** (see Section 4.3.6), the experimenter registers the question and the corresponding answer in the catalogue; this way, if another subject asks the same question later, the same answer can be given.

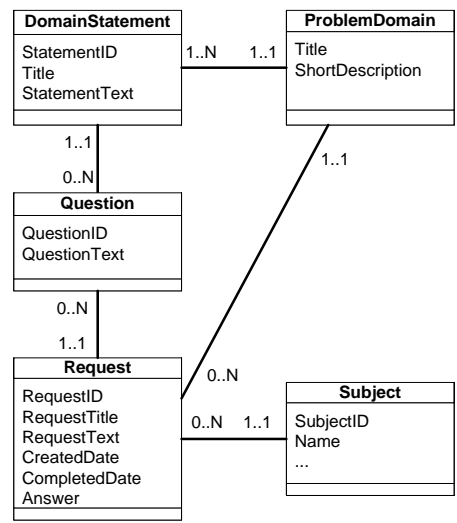

Figure 18.  Conceptual model of requirements elicitation in an experimental setting

In an experiment such as this one, in which requirements elicitation takes place, it is important that the elicitation process is enacted as repeateble as possible. This means that, if two subjects ask the same question (even if it is worded differently), they receive the same answer. Otherwise, a confounding factor is introduced (one subject could receive more information that the others) and there is even a threat to validity named *experimenter bias* (e.g. the experimenter could favour the subjects of a specific treatment group).

The help-desk elicitation intends to mitigate this issues. It offers the following advantages:

- The elicitation is asynchronous (as opposed to a face-to-face interview or a messenger-like interaction, which is synchronous), what gives the experimenter time the analyse the questions and plan the responses properly.

- If the experimenter is efficient, the elicitation becomes fluent and resembles a synchronous discussion.
- The helpdesk serves as a repository of requests and, together with the cataloque of answers-questions, facilitates the pre-RS traceability[8] and, therefore, the accountability of the experimenter's decisions.

The text of the request (RequestText) is writen by the subject and placed via the helpdesk; the helpdesk should provide the RequestID and the PlacementDate. Then the experimenter analyses the request and fragments its text into one or several questions, according to the question-answer catalogue. Figure 19 shows a how a request (R70) is split into two questions (Q85 and Q86) that refer to two specific statements (Sta53 and Sta61, respectively). Then the experimenter builds the answer to the request by means of combining the text of the statements and adding some elements that weave the conversation (introductory sentences with a phatic function and sentence connectors to weave different answers can be used, as long as they convey no extra information about the domain). The example with real data is shown in Section 5.8.

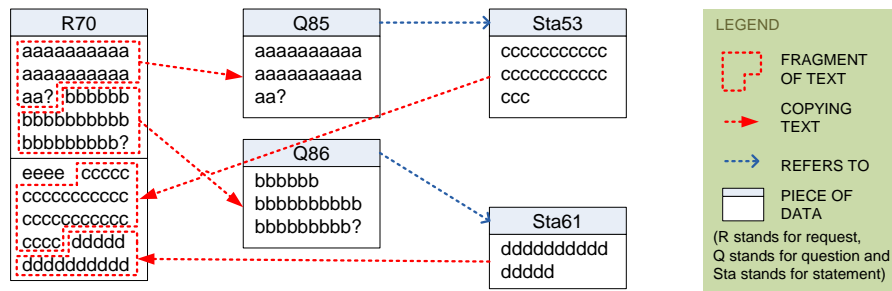

Figure 19. Fragmentation of requests into questions and composition of answers from statements

As the subjects discover information about the domain, they create a requirements model and the OO-Method conceptual model (the requirements model is optional in case they have not been trained in Communication Analysis yet). When the time alloted for the experimental task ends, the subjects hand in the conceptual model and (if any) the requirements model.

In order to capture the subjects' first impressions after using the respective methods to develop the models of the information systems corresponding to problems 1, 2 and 3, they responded a questionnaire: the **problem MAM survey** (see Section 4.3.7). The collected data is later used to evaluate the perception-based variables.

During the problem resolution, the subjects can express their doubts about the methods and their concerns about the experiment using a specific request category that is is defined in the helpdesk. The researchers only attend online those concerns that are vital for the continuation of the experiment (e.g. case-tool crashes); the doubts about the method are solved and discussed during the doubt-solving sessions.

## 3.4. Allocation of subjects to groups

After Problem 1, the subjects are split into two groups. Group A will next solve Problem 2 applying the OO-Method, while Group B receives training in Communication Analysis.

Subjects are allocated to groups randomly but the following restrictions are imposed:

---

[8] By pre-requirements specification traceability, Gotel et al. [1994] refer to those aspects of a requirement's life prior to its inclusion in the requirements specification; for instance, which user has expressed a given requirement, and in which interview.

- Blocking. Subjects that have a high knowledge of OO-Method (if any) are evenly distributed throughout both groups.
- Balance. Both groups should have (approximately) the same amount of subjects.

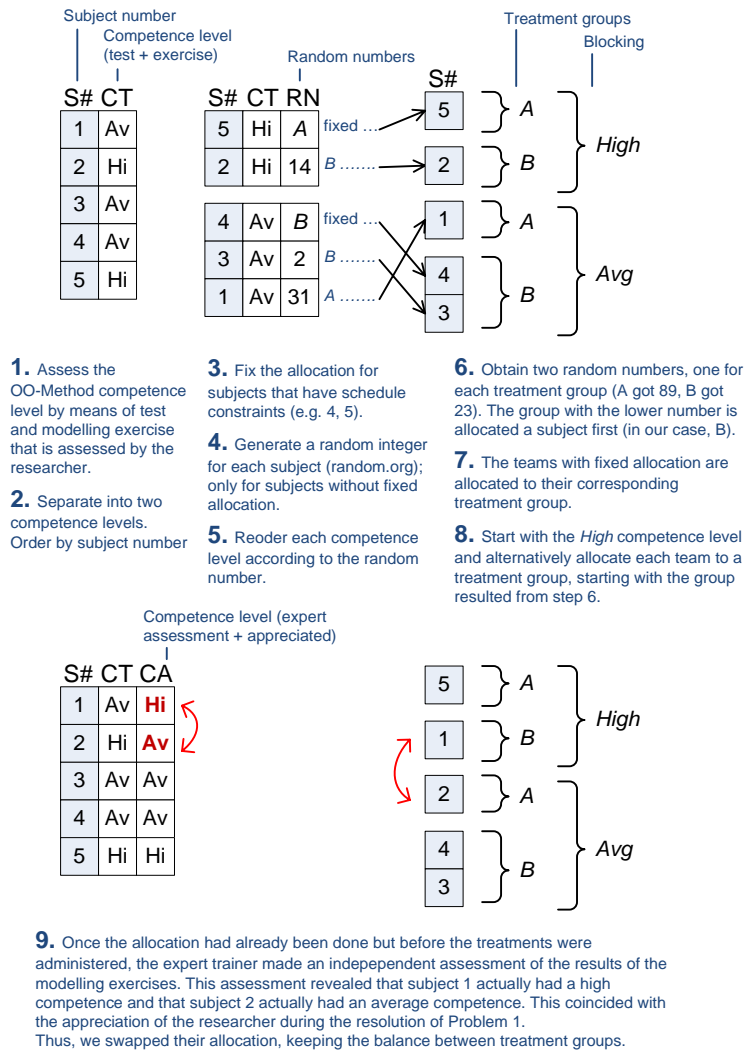

Figure 20.   Allocation procedure

Figure 20 shows the procedure we followed to allocate the subjects to the treatment groups.

## 3.5. Training on Communication Analysis

The subjects of Group A and Group B are trained on Communication Analysis separately. The **Communication Analysis training material** (see Section 4.4.1) covers the contents of the course[5], whose outline is presented in Table 5.

| Seminar # | Content |
|---|---|
| 1 | Requirements Engineering for information systems<br>Introduction to Communication Analysis<br>Communicative Event Diagram<br>Delivery of Communicative Event Diagram exercise (homework) |

| | | |
|---|---|---|
| 2 | Discussion on Communicative Event Diagram exercise | |
| | Communicative event specification | |
| | Delivery of event specification exercise (homework) | |
| 3 | Discussion on communicative event specification exercise | |
| | Requirements elicitation and analysis | |
| | Delivery of requirements elicitation and analysis exercise (homework) | |
| 4 | Discussion on requirements elicitation and analysis exercise | |
| | From requirements specification to conceptual models | |
| | Conceptual modelling exercise: Derivation of conceptual models | |
| | Final discussion on Communication Analysis | |

Table 5. Outline of the Communication Analysis training

During the course, a representative example of an Information System requirements model is developed. The subjects can use both the course material and the representative example as references during the experimental task.

The course includes training in a diagramming tool (e.g. Visio) that the subjects can use during the experimental task[9].

After the Communication Analysis training, the knowledge[10] of the subjects is assessed by means of a **Communication Analysis knowledge test** (see Section 4.4.2).

## 3.6. Wrap-up

This is the last seminar of the course. The reference solutions to Problems 1, 2 and 3 are discussed. The subjects are given a lecture on code generation technologies and a live demo of code generation is performed.

Standard ethical issues in experimentation should also be addressed in this seminar (e.g. subjects should be explained their contribution to the scientific community, how their data will be treated, etc.)

## 3.7. Conceptual model quality assessment

The assessment is carried out by one or several researchers using the list of statements (see Section 4.5.2), the list of questions (see Section 4.5.3) and/or the evaluation template (see Section 4.5.1). A detailed protocol is still to be defined. During the pilot experiment, only one researcher has acted as evaluator and he has used the list of statements.

---

[9] The diagramming tool enhances the comprehensibility of the Communication Analysis models by ensuring neat shapes and fonts.

[10] A modelling task would have been useful to assess their modelling competence, but we failed to prepare it on time.

# 4. INSTRUMENTATION

## 4.1. Kick off

### 4.1.1. In-take survey

We reviewed several online survey management sites (they reduce some threats and save time). The two final candidates were the following. They provided different features as part of the free account.

- SurveyGizmo: unlimited questionaires, maximum of 250 responses/month
- QuestionPro: 2 questionnaires at most, unlimited responses

Given the characteristics of this experiment (several questionnaires with at most 30 responses each), we opted for SurveyGizmo. However, in the following, the surveys are presented in textual form.

With regards to the in-take survey, in questions 1 to 3, only one answer per row is allowed.

Link to survey: http://www.surveygizmo.com/s/157014/in-take-survey

---

Please answer the following survey. It will help us know the background of the students of this course, so we can provide you a better training. It will also be useful for our research. Your answers will not influence your course grades.

**Previous knowledge**

Name and surname:

1. Rate your knowledge level of the syntax of the following information system analysis techniques.

1=Low; 2=Moderately low; 3=Average; 4=Moderately high; 5=High

|  | 1 | 2 | 3 | 4 | 5 |
|---|---|---|---|---|---|
| Data Flow Diagram |  |  |  |  |  |
| Entity Relationship Diagram |  |  |  |  |  |
| Activity Diagram |  |  |  |  |  |
| Class Diagram |  |  |  |  |  |
| Use Case Diagram |  |  |  |  |  |
| State Transition Diagram |  |  |  |  |  |
| Business Process Modeling Notation |  |  |  |  |  |
| Relational (database) Model |  |  |  |  |  |

2. Rate your experience in applying each of the following techniques and methods.

1=I have never user this technique; 2=I have seen examples in class; 3=I have solved small exercises; 4= I have solved moderately complex cases; 5=I have solved real cases professionally

|  | 1 | 2 | 3 | 4 | 5 |
|---|---|---|---|---|---|
| Data Flow Diagram |  |  |  |  |  |
| Entity Relationship Diagram |  |  |  |  |  |
| Activity Diagram |  |  |  |  |  |
| Class Diagram                       |   |   |   |   |   |
| Use Case Diagram                    |   |   |   |   |   |
| State Transition Diagram            |   |   |   |   |   |
| Business Process Modeling Notation  |   |   |   |   |   |
| Relational (database) Model         |   |   |   |   |   |

3. Assuming the following techniques would be available on your job, predict whether you would use them on a regular basis to analyse and design information systems.

1= Extremely unlikely ; 2= Quite unlikely; 3= Slightly unlikely; 4= Neither; 5= Slightly likely; 6=Quite likely; 7=Extremely likely

|                                     | 1 | 2 | 3 | 4 | 5 | 6 | 7 |
|-------------------------------------|---|---|---|---|---|---|---|
| Data Flow Diagram                   |   |   |   |   |   |   |   |
| Entity Relationship Diagram         |   |   |   |   |   |   |   |
| Activity Diagram                    |   |   |   |   |   |   |   |
| Class Diagram                       |   |   |   |   |   |   |   |
| Use Case Diagram                    |   |   |   |   |   |   |   |
| State Transition Diagram            |   |   |   |   |   |   |   |
| Business Process Modeling Notation  |   |   |   |   |   |   |   |
| Relational (database) Model         |   |   |   |   |   |   |   |

**Data modelling exercise**

> Make a data model in a paper sheet given the following description. Use the data modelling technique of your choice, but include a small legend with the notation conventions. Do not invest more than 10 minutes in this exercise.

4. Several bars that sell natural fruit juices have created The Fruit-Juice Club. There is a common catalogue of juices; each juice has a catchy name and it is made of one or more kinds of fruit (not more than five kinds of fruit per juice). Each bar sells some of these juices. People can become a member of the club in order to benefit from 10% discount in any of these bars. Adult members are given a plastic member card with their membership number, their name and their favourite fruit (they will have a double discount on those juices that contain this fruit). Children members get to choose two favourite fruits and they are given a sticker with each juice that includes any of these fruits. Adult club members are sent by postal mail one drink-card each month; the drink-card includes their name and their membership number, the current month and year, and twelve boxes that correspond to twelve drinks. Each time members order a drink, they can give the drink-card to the waiter; and then the waiter writes the name of the juice in the first empty box and stamps it. Each bar has its own stamp. If a member gets all the boxes stamped, s/he has been in three different bars that month, and s/he has ordered at least five different juices, then s/he is given a small prize. In addition, The Fruit-Juice Club wants to keep track of the total amount of drinks that a member has asked for (since they became members).

**Process modelling exercise**

> Make a process model in a paper sheet given the following description. Use the process modelling technique of your choice, but include a small legend with the notation conventions. Do not invest more than 10 minutes in this exercise.

5. A middleman company acts as intermediary between clients and suppliers. Most clients call the Sales Department, where they are attended by a salesman. Then the client requests one or several products that are to be sent to one or many destinations. The salesman takes note of the order. Other clients place orders by email or by fax. Then the Sales Manager reviews the order and assigns it to one of the many suppliers that work with the company. An order form is sent by fax to the supplier. The supplier receives the order form and checks whether they have enough stock or not. In case they have enough stock, they accept the order; otherwise, they reject it. In case the order is rejected, the Sales Manager assigns it to a different supplier (this can happen many times until the order is accepted). Once the order is accepted, the salesman sends a copy of the order to the Transport Department and the Insurance Department. In the Transport Department, the Transport Manager arranges how the goods will be carried to the destinations; he gives the logistics information to his assistant and the assistant sends it to both the client and the supplier. In the Insurance Department, the clerk specifies the insurance clauses, attaches them to the order form and sends the order form back to the Sales Department, where the salesman faxes the insurance information to the client. When the transportation vehicle (usually a truck, but sometimes a van) picks up the goods from the supplier's warehouse, the supplier phones the company to report that the shipments are on their way to their destinations.

The diagrams in Figures 21 and 22 are possible solutions to the modelling exercises.

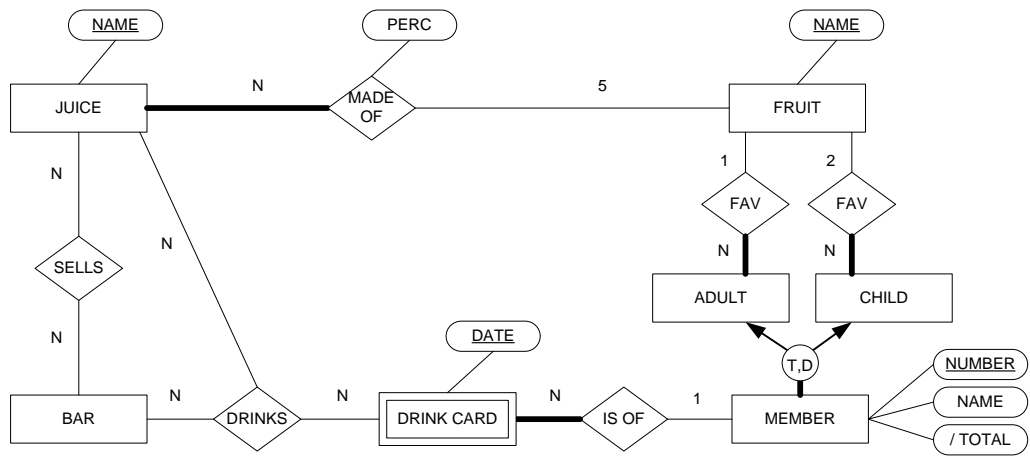

Restrictions:
- The sum of the percentages of fruit per juice shall amount to 100.
  SUM(Juice.made_of.perc)=100
- A member can only drink a juice in a bar if the bar actually sells the juice
  Bar.drinks.Juice IN Bar.sells.Juice
- A drink-card only has 12 boxes
  count(Drink_card.drinks)=12

Derived attributes:
- Member.Total=count(Member.is_of.Drink_card.drinks)

Figure 21. A solution to the data modelling exercise

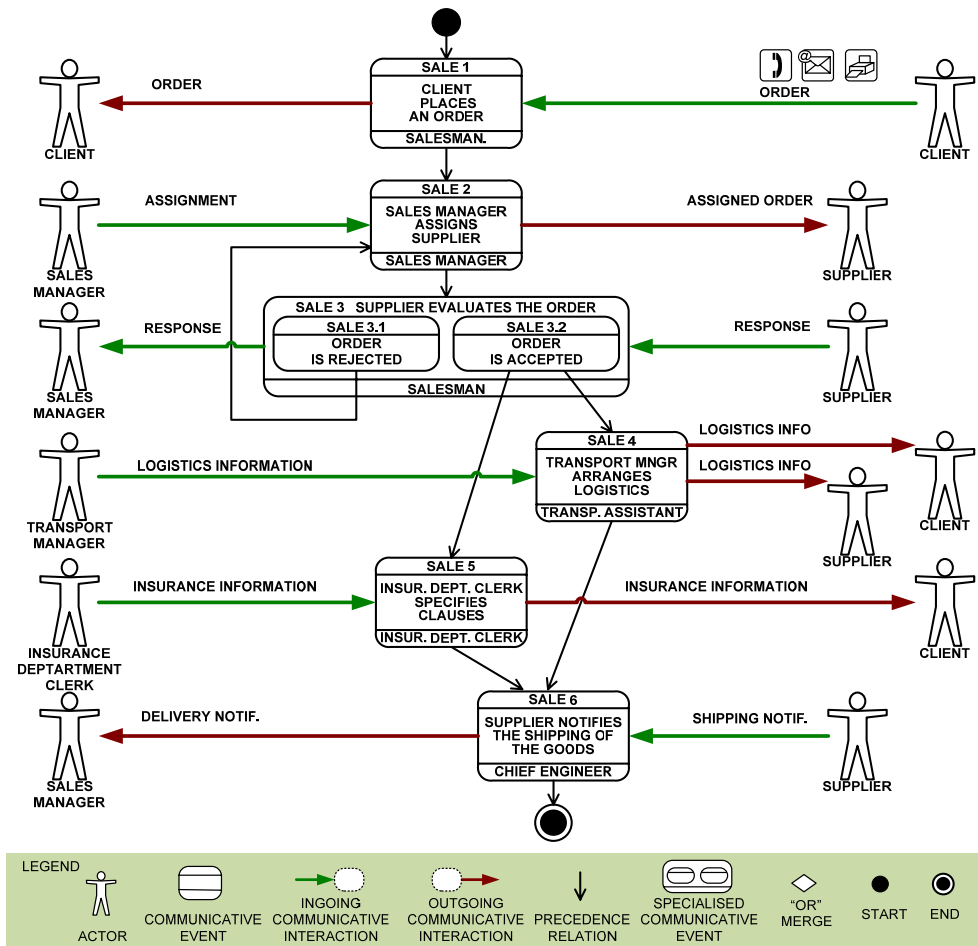

Figure 22.  A solution to the process modelling exercise

## 4.2. Training on OO-Method / OLIVANOVA

### 4.2.1. OLIVANOVA training material

The official training material is used. This includes slides that cover the course content, extended textual documentation that ellaborates on each of the conceptual model views, and exercises to put concepts in practice. The exercises are intended to incrementally build a conceptual model about an expense management application that later serves as a reference example.

Figure 23.  Example of the training material and snapshot of the *OLIVA**NOVA*** Modeler

The *OLIVA**NOVA*** Modeler is used to create the conceptual model.

## 4.2.2. *OLIVANOVA* knowledge test

The test consists of 30 multiple-choice questions.

http://www.surveygizmo.com/s/175633/olivanova-assessment

---

*OLIVA**NOVA*** knowledge level assessment

Please answer the following test about *OLIVA**NOVA***.
Name and surname

1. What does the Functional Model specify?
a)   The interface of the system.
b)   The static customer requirements.
c)   The changes in the value of the attributes of an object.
d)   None of the previous answers is correct.

2. Where are inheritance relationships defined?
a)   In the Object Model.
b)   In the Functional Model.
c)   In the Dynamic Model.
d)   In the Presentation Model.

3. What is the "Extended Creation" check-box (in the class-creation dialog) used for?
a)   If it is selected, *OLIVA**NOVA*** Modeler will show the 'Attributes' and 'Services' tabs.
b)   If it is selected, *OLIVA**NOVA*** Modeler will automatically add the creation, deletion and edition events to the new class and also the class identifier.
c)   If it is selected, *OLIVA**NOVA*** Modeler will automatically add the creation and deletion events to the new class and also a default State Transition Diagram for the class.
d)   None of the previous answers is right.

4. While defining a new class attribute, there is check-box named "Request upon creation"; what is it for?
(Note that creation time refers to the creation of a new object at run-time)
a)   If it is selected, a value for the attribute will be requested to the user at creation time.
b)   If it is selected, the attribute will take its value at creation time and this value will never change.
c)   If it is selected, the attribute is assigned the "default value" at creation time.
d)   None of the previous answers is correct.

5. What kind of attributes can we use as a class identifier?
a)   A variable attribute.
b)   A constant attribute.
c)   A null attribute.
d)   A derived attribute.

6. Which special check-box does a variable attribute have, that none of the other kinds of attributes have?
a)   Null.
b)   Keep Temporary Information.

c) Request Upon Creation.
d) Add to Edit Event.

7. Which statement describes how derivation rules work?
a) In order to use the derivation formula, all conditions must be fulfilled.
b) Conditions are checked top-down; if one condition is fulfilled, its corresponding derivation formula is used; if not, the next condition is checked.
c) All conditions are checked top-down; the derivation formula that is used corresponds to the last condition that is fulfilled.
d) None of the previous answers is correct.

8. How many derivation formulas can be defined for a derived attribute?
a) 0 or 1.
b) 0 or many.
c) 1 or many.
d) None of the previous answers is correct.

9. How many types of services are defined in OLIVA*NOVA*?
a) Three: Creation, Destruction and Edition.
b) Four: Creation, Destruction, Edition and Shared Events.
c) Two: Simple services and Complex Services.
d) Three: Events, Transactions and Operations.

10. What feature distinguishes Transactions from Operations?
a) The sequence operator.
b) The outbound arguments.
c) The all-or-nothing policy.
d) All of the previous ones.

11. What happens to a service when it is set as "Internal service"?
a) Nothing special.
b) It loses all its agent relationships and its links with the user interface.
c) It is not possible to invoke the service from transaction formulas of other classes.
d) None of the previous answers is correct.

12. Where can Outbound Arguments be used?
a) In a transaction formula, in order to initialize the values of inbound arguments of the subsequent services of the formula.
b) In the valuation rules of an event, in order to give value to attributes of the class.
c) In the constraints of the class.
d) None of the previous answers is correct.

13. When does an Outbound Argument take its value?
a) After checking the preconditions of a service.
b) At the beginning of the execution of the service.
c) After the execution of the service.
d) None of the previous answers is correct.

14. Which elements can be used in the precondition of a creation service?
a) Only constants and attributes of the class.
b) Constants, attributes and any rolepath of the class.
c) Constants, attributes and any operator.
d) None of the previous answers is correct.

15. What does the term "THIS" mean in a transaction formula?
a) The name of the class where the term "THIS" is used.
b) The current object in which the service is executed.

c) It can only be used in shared events, in order to represent the object that is the entry point for the event.
d) None of the previous answers is correct.

16. When does a integrity constraint have to be checked (at run-time)?
a) Only after finishing the execution of the creation service of the class in which the constraint is defined.
b) After finishing the execution of any service of the class in which the constraint is defined.
c) Before the execution of a service of the class in which the constraint is defined.
d) None of the previous answers is correct.

17. For which elements of a class can agent permissions be defined?
a) For attributes only.
b) For attributes and services only.
c) For attributes, events, and roles.
d) None of the previous answers is correct.

18. Can an agent class be agent of itself?
a) Yes.
b) No.
c) Only if the agent class is an administrator.
d) Only if the check-box "Self agent" of the class is marked.

19. Where can the formula "AGENT = THIS" be defined?
a) In preconditions of agent classes.
b) In Horizontal Visibility formulas, when the server class and the agent class are the same.
c) In no formula, since it is not a valid formula.
d) None of the previous answers is correct.

20. What is a precondition?
a) A condition than must be accomplished when the creation event is executed.
b) A condition that must be accomplished in all the states of an object.
c) A condition that must be accomplished before the service where it is defined is executed.
d) None of the previous answers is correct.

21. In an Aggregation relationship...
a) The aggregation relationship must be symmetric due to the Static property.
b) The minimum cardinality for the role is 2.
c) The visibility of attributes and services is bidirectional.
d) All the previous answers are correct.

22. With regards to roles...
a) From each class that takes part in the relationship, the class at the other end of the relationship is seen as a Role.
b) When two classes are connected by several relationships, the only way to differentiate one relationship from the others is by using Roles.
c) Roles are used in formulas to access to attributes and services of the related classes.
d) All the previous answers are correct.

23. With regards to dynamic relationships...
a) A relationship is dynamic if at least one role is dynamic.
b) Every dynamic relationship requires insertion and deletion shared-events.
c) A relationship is dynamic depending on the cardinalities of the roles.
d) None the previous answers is correct.

24. Does a "child" class have visibility over the attributes and services of its "parent" classes?
a) Yes, always.

b) A "child" class only has visibility over the direct "parent" class (i.e. the one immediately above in the specialisation hierarchy).
c) Yes, but only in the case that it has not been defined an attribute/service in the "child" class that has the same name as an attribute/service of the "parent" class.
d) None the previous answers is correct.

25. Does a "parent" class have visibility over the attributes and services of the "child" classes?
a) Yes, always.
b) A "parent" class only has visibility over the direct "child" class (i.e. the one immediately below in the specialisation hierarchy).
c) Yes, but only in the case that it has not been defined an attribute/service in the "child" class that has the same name as an attribute/service of the "parent" class.
d) None the previous answers is correct.

26. In which kind of formulas can user functions be used?
a) In transaction guards.
b) In initialisations of arguments.
c) In conditions of valuations and in constraints.
d) All of the previous answers are correct.

27. Can a service be executed if no transition has been defined for this service in the current state of the object?
a) Yes, it is possible as long as it is an internal service.
b) No, it is not allowed and an error message will be displayed.
c) Yes, it is possible if there is a transition in the STD that triggers this service (even if it is not defined in the current state of the object).
d) Yes, it is allowed if the corresponding button in the user interface has been added.

28. In a given class, a valuation is defined for each pair of...:
a) ...Events of the given class and attributes of the current class and the classes related by a rolepath.
b) ...Attributes of the given class and events of the current class and the classes related by a rolepath.
c) ...Events and attributes of the given class.
d) None of the previous answers is correct.

29. In an event/attribute pair that defines several valuation rules...
a) At least one valuation rule with an empty condition must be defined.
b) It is not necessary to define a valuation rule with a empty condition.
c) It is mandatory that all valuation rules define a condition.
d) None of the previous answers is correct.

30. The initial values for Atr1 and Atr2 are 2 and 3 respectively. The value of the arg_Value inbound argument is 4.
The modify event is executed; it defines the following valuation rules:

| Attribute | Event | Effect | | Condition | Current value |
|---|---|---|---|---|---|
| Atr1 | modify | = | arg_Value | arg_Value > 3 | |
| Atr2 | modify | + | Atr1 | | |

What will the final values for attributes Atr1 and Atr2 be?
a) Atr1 = 4, Atr2 = 7
b) These valuations will not be applied.
c) Atr1 = 7, Atr2 = 4
d) None of previous (write the values for Atr1 and Atr2 separated by a comma):

---

To obtain the list of correct responses please write an email to the corresponding author.

## 4.2.3. OLIVANOVA modelling exercise

The modelling exercise consists of a simple textual problem statement that describes a the management of shipment containers by a stevedoring company. The description is object-oriented to facilitate its translation into an *OLIVANOVA* conceptual model.

<div style="text-align: center"><b>How to manage the traffic of containers</b></div>

This document explains a study case about how to manage containers in a shipping company that owns a fleet of ships.

Relevant information about Ports:
- Port Code (4 characters)
- Name (String - Mandatory and editable)
- City (String - Optional and editable)
- Maximum number of moorings (Natural - Mandatory and editable)
- Number of current ships moored in the Port

Relevant information about Ships:
- Ship Code (4 characters)
- Name (String - Mandatory and editable)
- Maximum weight supported (Real - Mandatory and editable)
- Current weight of the containers embarked.

Relevant information about Containers:
- Identifier number (Integer autonumeric)
- Weight (Real – Mandatory and editable)
- Is Delivered? (Boolean – Mandatory and editable)

Relevant information about Port Controller:
- Identifier number (Integer autonumeric)
- Name (String – Mandatory and editable)

Relevant information about Administrator:
- Identifier number (Integer autonumeric)
- Name (String – Mandatory and editable)

When a container is created, it is placed in a Port defined as current Port. Besides, a container must indicate which is its Target Port and its weight. This weight can be modified if the container is placed in Port, never when it has been embarked in a ship.

A container can be embarked onto a ship if it is moored in the same Port where the container is placed. Each ship has specific weight limit. It will not be possible to embark a container if the maximum weight of the ship is reached.

A container can be disembarked in the Port where the Ship is moored. A container is considered delivered, when it is disembarked in the Port and it is equal to the Target Port. When a container is delivered, it cannot be embarked onto any ship and it can be removed from the system.

Port Controllers are the responsible to embark and disembark containers onto the ships in a specific Port. They can manage only the containers of their port and the containers of the ships moored in it. Besides, they allow a ship to moor in their port. They can see all ships except those ships that are moored in another port. Additionally, in order to prevent problems, there are administrators that can manage anything in the application.

There exists a reference solution provided by CARE Technologies. Just as an example, Figures 24 and 25 show the Object Model and the services of the class Container respectively.

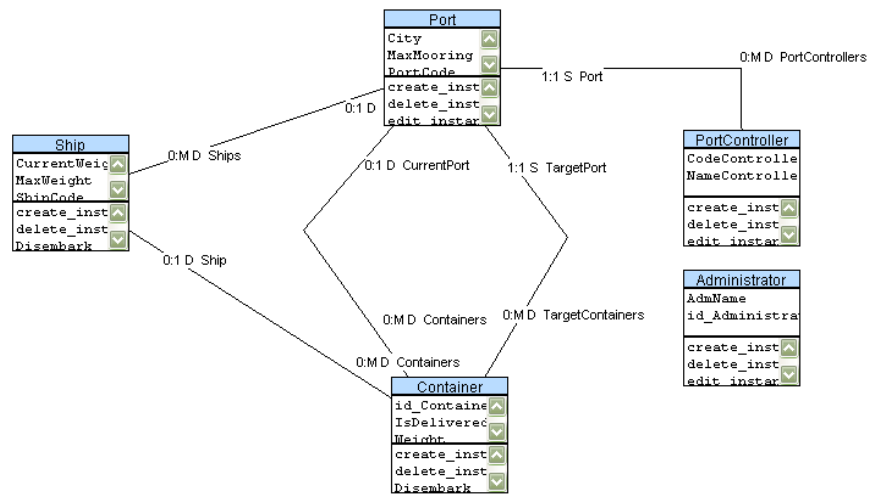

Figure 24. An Object Model that corresponds to the Containers problem statement

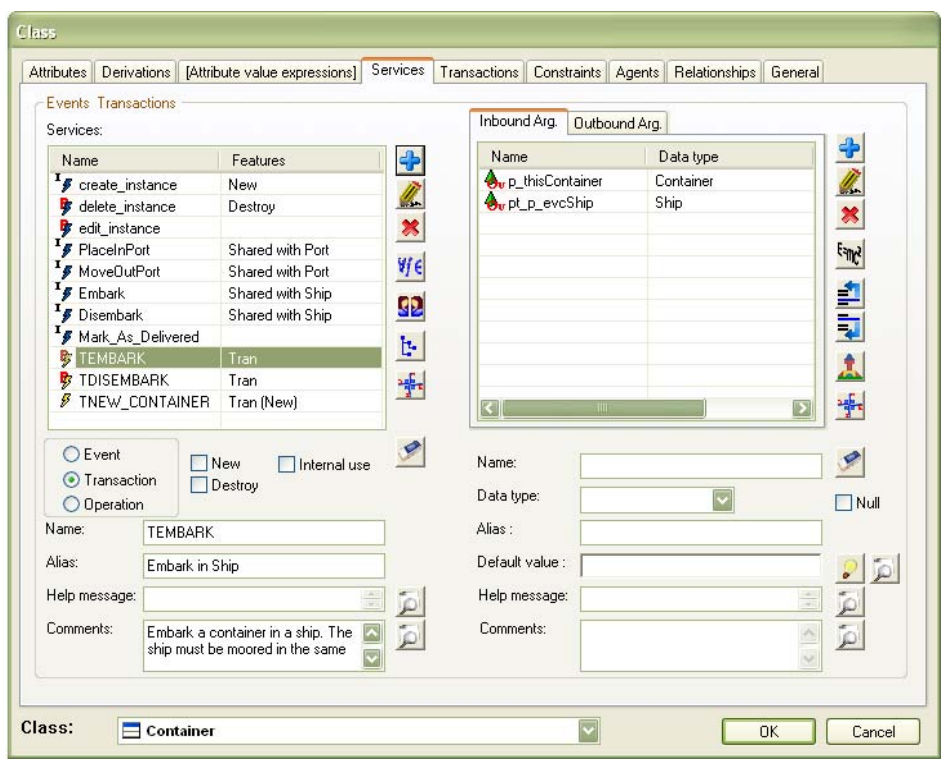

Figure 25. Services of the class Container

### 4.2.4. *OLIVANOVA* modelling exercise correction template

The template consists of a list of statements about the problem domain. The evaluator has to inspect the conceptual model so as to assess whether the substatements are included in the model or not. Each substatement has hints for the evaluator (in order to facilitate the conceptual model inspection).

The correction template is derived from the requirements specification of the modelling exercise (see Figure 32).

Figure 26. Partial view of the correction template for the modelling exercise

## 4.3. Resolution of Problems 1,2,3

### 4.3.1. Helpdesk system

Several helpdesk systems were investigated; we chose ManageEngine ServiceDesk because it is web-based and it offers a free license for evaluation purposes (although it is limited to only one technician, this is not a problem in our experimental setting the experiment: we had just one researcher answering the subject's questions). The helpdesk was installed in a server in University of Twente. The following configurations were made via the configuration wizard:

- **Categories**. When users of a helpdesk place a request, they categorise it according to a predefined category, in order to facilitate allocating the request to the appropriate technician. We used categories to allow the subjects indicate the context of each request; that is, the experimental task the request is related to. Table 6 shows the categories that were created.

| Category | Description intended for the subjects |
| --- | --- |
| Problem 1 requirements elicitation | Use this category in order to elicit requirements of Problem 1 (Projects office) |
| Problem 2 requirements elicitation | Use this category in order to elicit requirements of Problem 2 (Photography Agency) |
| Problem 3 requirements elicitation | Use this category in order to elicit requirements of Problem 3 (TUO master management) |
| OLIVANOVA methodological doubt | Use this category in order to place questions and doubts concerning the method (*OLIVANOVA*) |

Table 6.  Helpdesk categories used in the experiment

- **Request template**. Each request conforms to a given template (a set of data fields arranged in a web form). We restricted this template to the minimum that was necessary for the experiment. Figure shows the configuration of such template.

Figure 27.  The default template for requests

- **Requesters**. Each user of the helpdesk needs to have its own account. We created one account for each experimental subject. Each requester record was filled with the name of the subject, a login consisting of their surname, and an initial password that they had to change. The field "Requester allowed to view" was set to "Show only their own requests"[11].

- **Technicians**. The free account only allows for one technician. The administrator account was the only technician we defined.

---

[11] This is important in order to prevent the experimental subjects to see other subject's questions, which would introduce a confounding factor.

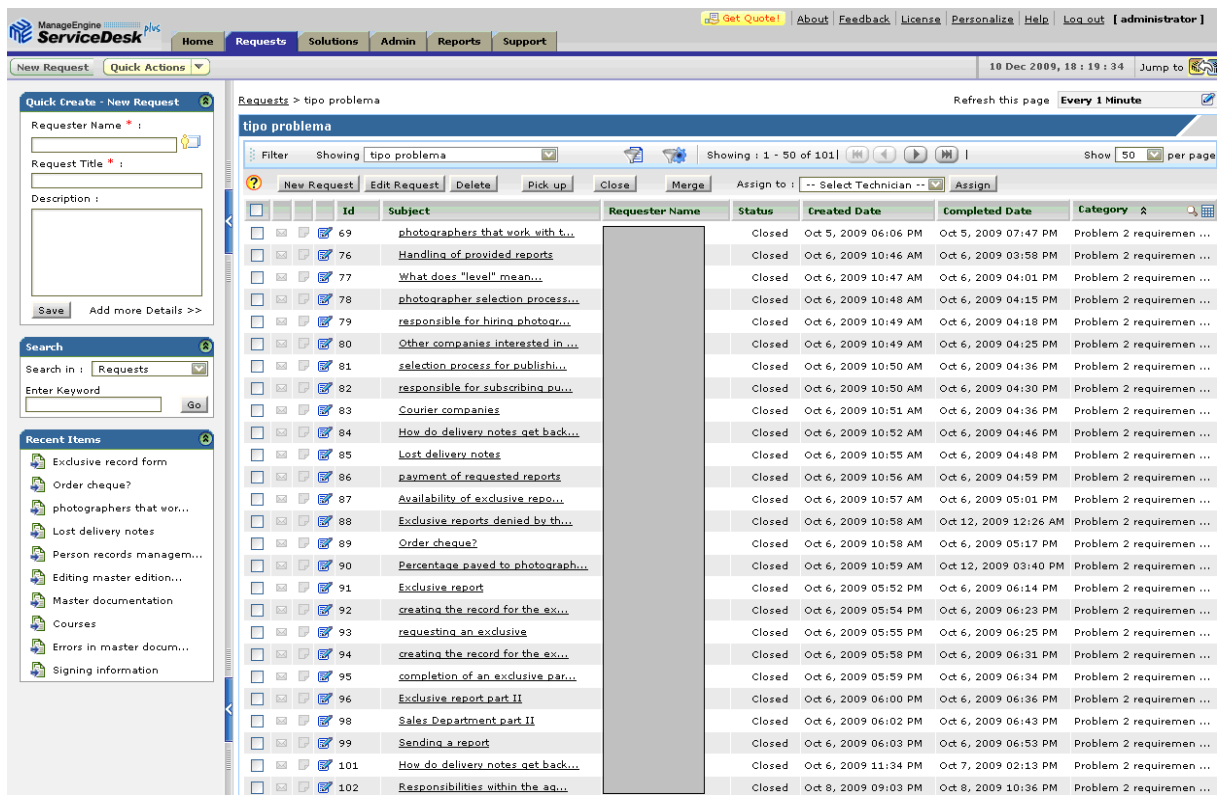

Figure 28. Snapshot of the servicedesk list of requests[12]

## 4.3.2. Experimental task instructions

The task instructions are similar for the three experimental tasks. The subjects are informed of the objective of the task, a short explanation on how to use the helpdesk for requirements elicitation is provided, and the due date is fixed. In the following, the task instructions for Problem 1 are provided[13].

**Task instructions. Problem 1**

**Objective of the task**
The objective is **to make the *OLIVANOVA* conceptual model** that corresponds to the information system that a certain organization needs.
So, what you have to hand in at the end of the task is the file that contains your conceptual model; that is, the .OOM file.
You have been given a short problem statement. This introduces the information system that you have to model, but it does not give many details about it. You need to ask for the rest of the requirements. Note that this differs from what you have done in class until now (you were given a complete textual specification of the system, in a language that was very close to the object-oriented paradigm.
**How to elicit requirements**

---

[12] The bames of the requesters have been omitted to keep the anonymity of the experimental subjects.

[13] The server name has been omitted.

In order to discover the details about the problem, you have to place questions in a helpdesk (a.k.a. servicedesk) that is available at the following URL:

    http://servername.ewi.utwente.nl:8080/

Note: I have tested that it works fine with Internet Explorer and Firefox. I cannot tell about other web browsers.

Each of you has a personal account as a 'requester' (those users that place questions). The initial login data is the following:

    *Username*: the last part of your surname
    *Password*: the last part of your surname

You have to change your password. To do this, login into the servicedesk, click on *Personalize* (top right), and a pop-up window will appear, where you have a *Change password* tab.

The only functionality you will be using is placing questions (a.k.a. requests) and reading them.

**Placing requests**

Just click on *New request* (in several places, e.g. green button top left) and a form will appear.

    *Requester details*: you do not need to fill any of these fields (*Name*, *Contact number*, *Department*) since you have logged in with your user account.
    *Category*: select one category, depending on the type of question (see explanation below)
    *Subject*: write a few words that will help (you and us) recognise the request in a list of requests.
    *Description*: write the question or inquiry here.

You can place three types of questions (you have two chose the corresponding Category).

    *Problem 1 requirements elicitation*: questions that are related to the problem domain; you place them in order to discover details about the way the organisation works, in order to create the appropriate conceptual model. Try to be concise and try to focus on one question per request. You will notice that our answers try to simulate those of a real user; so you cannot ask as if you were interviewing an *OLIVANOVA* expert (that is, do not ask about classes or transactions, otherwise, the users will not understand your question). This type of questions will be answered during the assignment.
    *OLIVANOVA methodological doubt*: questions that are related to the method; you place them whenever a doubt or a difficulty arises. In principle, these questions will not be answered during the assignment, but we will prepare a class for solving these doubts, so it is really worth for you to register them.
    *Other doubts, difficulties and concerns*: feel free to also place doubts and difficulties about requirements elicitation or any other subject that is not covered by the previous categories. If what you write suggests that you are expecting an answer, you will be answered during the assignment, otherwise, we can discuss it in class or we will just keep it for the record. Feel free to even place complaints; as we have told you before, we are not going to grade you for anything else but your performance on the assignments, and we are not going to judge you for anything else but your respect and your commitment. Furthermore, we are still available by email, phone or in person, if you find it more convenient.

You are not expected to use any requirements specification technique. But if you wish you can create a textual specification by compiling your questions and answers and organising them as you wish.

**How to create the conceptual model**

You have to use *OLIVANOVA* Modeler to create the conceptual model. The software has been uploaded to Blackboard. We have made it available in the section Course material; it is compressed as a ZIP file and it is a standalone package (which means that you do not need to install anything, just uncompress it and run the EXE file).

To create the conceptual model you should proceed as it has been explained and practiced during the training.

**Planning**

We expect the following deliveries:

*Snapshot delivery*: Tuesday 15, 23:59 (at the latest). You are expected to send us the model as it is. You will not be graded for this incomplete model, but we want to monitor your performance.

*Due date*: Thursday 17, 15:59 (at the latest). You are expected to send us the model. You will be graded for this model with regard to syntactic, semantic and pragmatic quality (in short, well formed, complete, understandable models will get good marks).

### 4.3.3. Short problem description

The short problem description only intends to offer an overview of the organisation and its work practice, but it does not provide all the requirements of the information system. In the following, the description of Problem 1 is shown.

---

A projects office carries out industrial electrical installations for their customers (the customers are typically companies). The projects office staff consists of a clerk, a chief engineer, an accountant, engineers and workers.

The usual work practice is the following. A company contacts the projects office clerk, who opens a project record where data about the requested project and the company is registered (e.g. company data, project description). The record is placed on the 'New Projects' tray of the chief engineer. Every morning, the chief engineer checks new projects and he assigns them to the most suitable engineer. The chief engineer returns the project record to the clerk, who stores it in a tray labelled 'Assigned projects'. The clerk also updates the assigned engineer's record by increasing the number of projects that the engineer is working on.

When an engineer finds out that s/he has been assigned a project, s/he gets in touch with the company's contact person and makes a date in order to visit the company. The assigned engineer visits the company and gathers the customer needs (by talking to the customer and inspecting the place). Back in the office, s/he draws up a budget and hands it to the clerk, who sends a copy of the budget to the company for approval. The engineer assigns the project to as many workers as needed and the project becomes active. The assigned engineer and the workers carry out their tasks and, once the work is done, the engineer reports this fact to the clerk, so that the accountant can manage the invoicing.

Two forms are attached as an example of project records[14]. The projects office currently manages information by means of paper forms (except for a commercial off-the-shelf invoicing application). The aim of this software development is to replace paper forms with a software application.

---

[14] The forms included in the original problem description were bigger, we have reduced them for reasons of space.

| PROJECT RECORD | | | | | PROJECT RECORD | | | |
|---|---|---|---|---|---|---|---|---|
| **PROJECT** P0034/08 | | **Requested** 24-04-2008 | | | **PROJECT** P0128/06 | | **Requested** 10-11-2006 | |
| **COMPANY:** 19.345.631-Q | VAT# | Company name Delicioso Olive Oil, SA | Address Poligono Sur prc.64, n.7 | | **COMPANY:** 44.409.106-T | VAT# | Company name HydraMax, SA | Address C/ Cervantes, n.56 46173 Alfabeguer de les Canyes |
| **Contact person:** Sergio Pastor González | | | | | **Contact person:** Unai Erviti Bowen | | | |
| **Contact telephone:** 96 3870000 ext. 83534 | | | | | **Contact telephone:** 555 675 910 | | | |
| **Description:** | Install air conditioning unit in a 1000m2 warehouse | | | | **Description:** | Solar panels on the roof and uninterruptible power supply for water supply facilities. | | |
| **ENGINEER** Assigned on 26-04-08 | Oscar España del Rio | | ID# 22.412.336-S | | **ENGINEER** Assigned on 11-11-06 | Gon Peim | | ID# 45.376.478-L |

*Project record budget tables for P0034/08 (Total: 15.573,00 €) and P0128/06 (Total: 83.720,50 €).*

### 4.3.4. Textual requirements models

For each problem, a textual requirements model is available. It describes the informational needs of the organisation in great detail, and it includes many scanned business forms. This model is not delivered to the subjects. The experimenters can use it as a reference specification of the problems to be modelled by the subjects. This instrument is not included herein for reasons of space.

### 4.3.5. Communication Analysis requirements models

For each problem, a Communication Analysiss requirements model is available. It consists of a description of the organisation, a communicative event diagram depicting the business process model that corresponds to part of the organisational work practice, and a set of event specification templates (one for each communicative event in the diagram). This instrument is not included herein for reasons of space, but the reader can refer to [España, González et al. 2011], where a reference example is shown.

### 4.3.6. Question-answer catalogue

The helpdesk system does not fully suit our needs concerning the experiment, we had to store the catalogues of questions in Microsoft Word documents. The catalogues were embedded into the Communication Analysis requirements models of the problems at hand; that is, statements and questions were inserted as tables within the requirements model, as close as possible to the section related to the statement (see Figure 29). This way, the document structure provided an index to the catalogue, thus facilitating search, retrieval and insertion of questions.

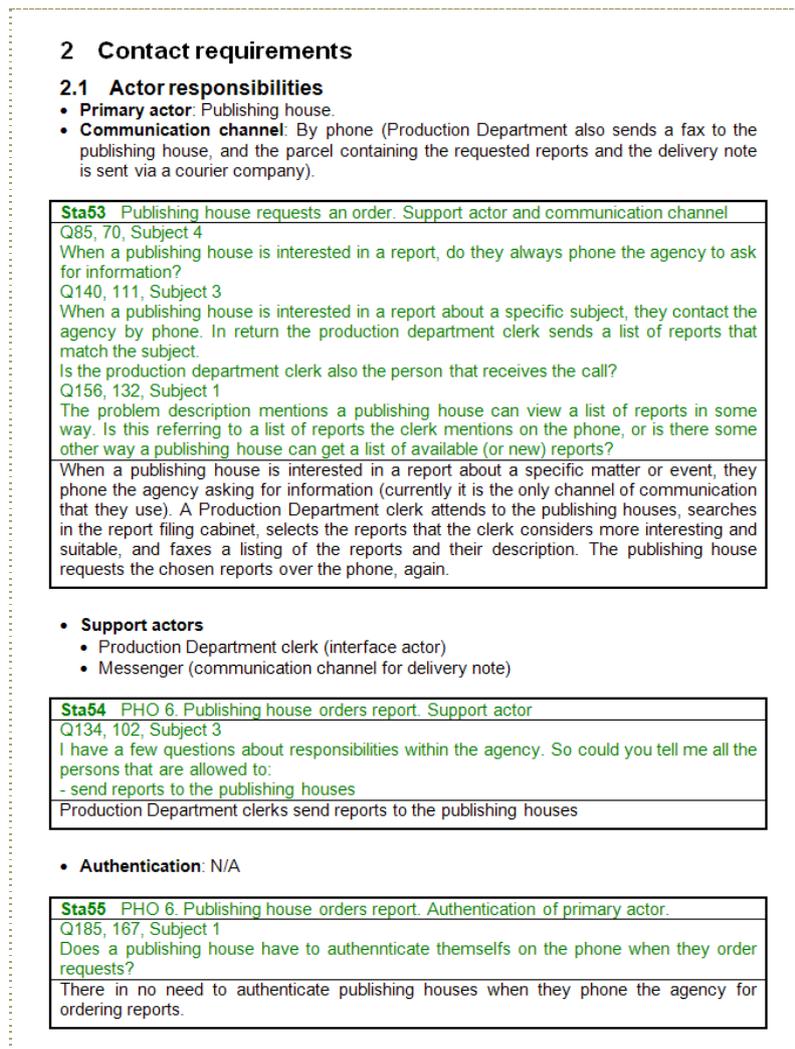

Figure 29.  Fragment of the requirements model of Problem 2 with the embedded catalogue of questions

The template that we used is the following.

| **StatementID**  Title |
|---|
| (All questions asking about the statement are included here.) |
| QuestionID[15], RequestID, SubjectID |
| QuestionText |
| StatementText |

### 4.3.7. Problem 1/2/3 MAM survey

The MAM survey is similar to the one used in [Moody 2003]; it has only been slightly adapted to suit the methods evaluated in this experiment. All the questions are rated by the experimental subject on a 7-point Likert scale.

Neither the question number (e.g. Q1) nor the reference to the original MAM survey item (e.g. PEOU1) appear in the survey that is presented to the subjects.

---

[15] The question identifier was actually omitted, but we include it in the Figure 29 to facilitate traceability with the examples included in Sections 3.3 and 5.8.

http://www.surveygizmo.com/s/157035/problem-1-survey
http://www.surveygizmo.com/s/157312/problem-2-survey-a
http://www.surveygizmo.com/s/157313/problem-2-survey-b
http://www.surveygizmo.com/s/157314/problem-3-survey

---

This survey is related to your perceptions of the method you have applied to analyse Problem 1; that is, the OO-Method.

Name and surname:

Rate the extent to which you agree with each statement.
The values of the 7-point scale correspond to:
      1 = Strongly disagree
      2 = Disagree
      3 = Disagree somewhat
      4 = Undecided
      5 = Agree somewhat
      6 = Agree
      7 = Strongly agree

| | 1 | 2 | 3 | 4 | 5 | 6 | 7 |
|---|---|---|---|---|---|---|---|
| Q1. I found the procedure for applying the method complex and difficult to follow (PEOU1) | | | | | | | |
| Q2. I believe that this method would reduce the effort required to analyse information systems (PU1) | | | | | | | |
| Q3. Information systems modelled using this method would be difficult for other analysts to understand (PU4) | | | | | | | |
| Q4. Overall, I found the method difficult to use (PEOU2) | | | | | | | |
| Q5. This method would make it easier for analysts to verify whether conceptual models are correct (PU4) | | | | | | | |
| Q6. I found the method easy to learn (PEOU3) | | | | | | | |
| Q7. Overall, I found the method to be useful (PU4) | | | | | | | |
| Q8. Using this method would make it difficult to discover the needs of the organisation (PU5) | | | | | | | |
| Q9. I found it difficult to apply the method to the information system I was trying to analyse (PEOU3) | | | | | | | |
| Q10. I would definitely not use this method to analyse information systems (ITU1) | | | | | | | |
| Q11. I found the method clear and easy to understand (PEOU5) | | | | | | | |
| Q12. Overall, I think this method does not provide an effective solution to the problem of analysing information systems (PU6) | | | | | | | |
| Q13. Using this method makes it easy to capture requirements (PU5) | | | | | | | |
| Q14. I am confident that I am now competent to apply this method in practice (PEOU6) | | | | | | | |
| Q15. Overall, I think this method is an improvement to other UML-based methods (PU4) | | | | | | | |
| Q16. I intend to use this method in preference to other analysis methods if I have to develop information system in the future (ITU1) | | | | | | | |

## 4.4. Training on Communication Analysis

### 4.4.1. Communication Analysis training material

The training material in English has been created based on available teaching material in Spanish from the course "Conceptual modelling of information systems" taught in Universitat Politècnica de Valencia (Spain). The material includes slides that cover the course content, extended textual documentation that ellaborates on each topic, and exercises to put concepts in practice. A reference example is developed throughout the sessions.

Figure 30. Example of the training material (slides and textual)

### 4.4.2. Communication Analysis knowledge test

A test with 24 multiple-choice questions is used for the Communication Analysis modelling competence assessment.

http://www.surveygizmo.com/s/190622/ca-assessment

**Communication Analysis knowledge level assessment**

Please answer the following test about Communication Analysis.
Name and surname

1. What is, under the point of view of Communication Analysis, the two aspects of reality that need to be described?
a) Actors and processes.
b) Communicative events and linked communications.
c) Static and dynamic aspects.
d) Dynamic and volatile.

2. What are unity criteria used for in Communication Analysis?

a) They are modelling primitives to draw communicative events.
b) They are norms that guide the encapsulation and identification of business processes.
c) They are guidelines for specifying communication structures.
d) They are techniques that allow the analyst to manage requirements traceability.

3. What three unity criteria does Communication Analysis propose?
a) Trigger unity, communication unity, reaction unity.
b) System, subsystem, process.
c) Precedence relationship, linked communication, logical connector.
d) None of the previous answers is correct.

4. What is a communicative interaction?
a) A communicative event of the information system.
b) An interaction between actors with the aim of exchanging information.
c) The interaction between an external actor and the organisation.
d) The usage of a software application by an actor with the intention to obtain some listing.

5. Which is eminently, the main direction of information in an outgoing communicative interaction?
a) There is no message conveyance in outgoing communicative interactions.
b) From the actor to the information system.
c) From the information system to the actor.
d) None of the previous answers is correct.

6. Can a communicative event involve information input?
a) No, it only involves information output.
b) Yes, but it is not necessary if the information is not meaningful.
c) Yes, it always involves an input of new meaningful information.
d) None of the previous answers is correct.

7. Can a communicative event involve information output?
a) No, it only involves information input.
b) Yes, but only in case that there is a support actor.
c) Yes, it can involve information output.
d) None of the previous answers is correct.

8. Can an outgoing communicative interaction involve information input?
a) No, it only involves information output such as listings and output forms.
b) Yes, but only in case that there is a primary actor that introduces a selection criterion.
c) Yes, for instance if the organisation wants to audit its occurrence.
d) None of the previous answers is correct.

9. Does a communicative event change the memory of the information system?
a) Yes, always.
b) Usually yes, but not always.
c) No, never.
d) None of the previous answers is correct.

10. Which sequence of actions better describes what a communicative event encapsulates?
a) Retrieval – processing – distribution – storage.
b) Retrieval – processing – distribution.
c) Acquisition – retrieval – processing – distribution.
d) Acquisition – validation – processing – storage.

11. Which definition of event corresponds better with the use of this term in Communication Analysis?
a) A stimulus that interrupts a system component activity.
b) An abstraction of a change of state in the organizational domain.

c) An individual stimulus from one object to another.
d) A stimulus that can trigger a state transition.

12. What kind of business process diagram is a Communicative Event Diagram?
a) It is a local business process diagram because it refers to one organisation.
b) It is a global business process diagram because it may involve several business objects.
c) It is a local business process diagram because it depicts the events that can happen to a specific business object.
d) It is a global business process diagram because it can depict a business to business interaction.

13. Given the following communicative event diagram, which statements are true?
(One or several statements can be true).

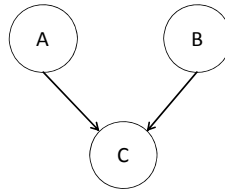

a) In order to occur C, either A or B have to occur first.
b) After A has occurred, C necessarily has to occur.
c) If A and B have occurred, C necessarily has to occur.
d) In order to occur C, both A and B have to occur first.

14. What kind of specialisation does the Communicative Event Diagram use for communicative events that have variants?
a) Static specialisation.
b) Dynamic specialisation.
c) Encapsulated specialisation.
d) Successive classes.

15. Which of the following statements does not correspond to the trigger unity criterion?
a) Trigger responsibility is external.
b) Some actor establishes contact with the IS and triggers organisational reaction.
c) The event occurs as a response to an internal interaction.
d) Each communication that is established with the IS is provoked by the occurrence of a unique phenomenon.

16. Which of the following statements is false?
a) A communicative event triggers the information system reaction, which is a composition of synchronous activities.
b) Two occurrences of two consecutive classes of normative events can never occur synchronously.
c) Two occurrences of the same class of communicative events can occur synchronously.
d) Two occurrences of the same class of communicative events can occur asynchronously.

17. Place the following requirements levels in ascending order, as Communication Analysis proposes them?
a) Process.
b) System/subsystems.
c) Communicative interaction.

18. Which of the following statements about actors is false?
a) The primary actor triggers the communicative event by establishing contact with the organisation and provides the conveyed input information.
b) The actor that is in charge of physically interacting with the information system interface in order to encode and edit input messages is considered a support actor.

c) The support actor of a communicative event does not new meaningful information to the information system.
d) Those actors that receive an outgoing communicative interaction are considered receiver actors, and they can be members of the organisation or external entities.

19. Which of the following types of requirements does not belong to the communicational content category?
a) Domains for each message field.
b) Derivation formulas for message fields.
c) Restrictions on the message structure.
d) All of the previous options do belong to the communicational content requirements category.

20. To which requirements category belongs a restriction on when an event can occur belong?
a) Contact requirements.
b) Communicational content.
c) Reaction.
d) It does not belong to any of the previous options.

21. In the communicative event specification template, linked behaviour refers to...
a) A description of the information system reaction; that is, what changes are made or recorded in the system.
b) Who needs to know something about this communicative event in order to make decisions or in order to act in a certain way.
c) Business rules o complex conditions that influence future reactions (to future events) in terms of the values of certain fields of the communication structure.
d) None of the previous answers is correct.

22. Select the statement about form analysis that is false.
a) Form analysis is part of generative analysis, which aims to discover and specify a first version of the business processes.
b) The analyst should first focus on input forms and then analyse output forms.
c) In form analysis, the analyst has to investigate who are the sources of the information with which the form is filled.
d) Form analysis allows creating one communication structure per form, in order to specify the message that corresponds to the communicative event.

23. Select the statement about forms that is false.
a) Data input forms are the starting point for form analysis.
b) Roundtrip forms are associated with only one communicative event, so it is important to identify them properly.
c) Intermediate results forms contain summarised (redundant) information and calculated fields in order to have useful information readily available.
d) Listing forms can be used for auditing the usage of the information system or to consult information that is necessary to carry out daily operations of the organisation.

24. Select the statement about revision of communicational behaviour that is false.
a) The analysis of exceptions and alternatives allows the analyst to discover exceptions in both internal and external treatments.
b) A conditioned registry is a communicative event that (conditionally) needs to be triggered before another communicative event, in order to register some information.
c) There are three types of linked communications: prior outputs, issuance audits and reception audits.
d) In a sequence of normative events, the users expect that, once a given event has occurred, the subsequent event should occur within a reasonable time frame.

To obtain the list of correct responses please write an email to the corresponding author.

## 4.5. Conceptual model quality assessment

### 4.5.1. Problem 1/2/3 quality assessment: Likert scales (RAT)

The following survey supports the assessment of model quality based on the framework by Lindland et al.

http://www.surveygizmo.com/s/174619/conceptual-model-quality-evaluation-likert

Dear evaluator. Please inspect the conceptual model that corresponds to the conceptual model identifier that we provided you and fill in this quality evaluation form. Make sure you note down the time before starting the evaluation and after you finish it. We ask for your name only for traceability purposes. Your name will not be made public.

Your name and surname:

Evaluation identifier code ("EV plus 3 digits", you can find this code in your assignment):

Conceptual model identifier code ("CM plus three digits", you can find this code in your assignment and it coincides with the name of the model file):

In order to assess the quality of the model, inspect it and then rate it according to the following goals (read the following explanation if you are not familiar with the conceptual model quality framework by Lindland, Sindre and Solvberg):

**Syntactic quality**

Syntactic quality is the degree to which the model adheres to the OO-Method/OLIVANOVA language rules. Syntactic errors and deviations from the rules decrease syntactic quality. There is only one syntactic quality goal: syntactic correctness.

> Syntactic correctness: The extent to which the statements (a.k.a. elements) of the model are according to the syntax. A model if correct if all the statements adhere to the language rules.

**Semantic quality**

Semantic quality is the degree to which the model represents the problem domain. The more similar the model and the domain, the better the semantic quality. There are two semantic quality goals: feasible validity and feasible completeness.

> Feasible validity: The extent to which the statements of the model are correct and relevant to the problem domain. Invalid statements are those that do not pertain to the problem or express something incorrectly. A model has achieved feasible validity when there is no invalid statement in the model, such that the additional benefit to the conceptual model from removing the invalid statement exceeds the drawbacks of removing it.

> Feasible completeness: The extent to which the model includes the relevant and correct statements about the problem domain. If there is a relevant statement about the domain that is not included in the model, then the model is incomplete. A model has achieved feasible completeness when there is no relevant statement about the domain, not yet included in the model, such that the additional benefit to the conceptual model from including the relevant statement exceeds the drawbacks of including it.

**Pragmatic quality**

Pragmatic quality is the degree to which the model is correctly interpreted by its audience. The less misunderstanding, the better the pragmatic quality. There is only one pragmatic quality goal: feasible comprehension.

> Feasible comprehension: The extent to which the model is understood by the people who read it; that is, the extent to which the model allows you (the reviewer) to understand it. Several aspects influence comprehension; for instance, the layout of model elements, the relevance of the names of model elements, the way the model is structured, the explanatory comments it contains.

Start time (the time when you start the evaluation; before inspecting the conceptual model):

Rate the model quality. The values of the 7-point scale correspond to:

   1 = Extremely low quality
   2 = Very low quality
   3 = Low quality
   4 = Medium quality
   5 = High quality
   6 = Very high quality
   7 = Extremely high quality

|  | 1 | 2 | 3 | 4 | 5 | 6 | 7 |
|---|---|---|---|---|---|---|---|
| Syntactic quality: Syntactic correctness |  |  |  |  |  |  |  |
| Semantic quality: Feasible validity |  |  |  |  |  |  |  |
| Semantic quality: Feasible completeness |  |  |  |  |  |  |  |
| Pragmatic quality: Feasible comprehension |  |  |  |  |  |  |  |

End time (the time when you finish the evaluation; after filling this form):

---

The template gives value to variables *Completeness_STA*, *Validity_STA* and *Comprehension_STA*. Syntactic correctness can be disregarded and even omitted from the template if *OLIVANOVA* Modeler has been used to create the conceptual models, since the Modeler itself can provide the number of syntactic errors of a model.

The Likert value selected by the reviewer is directly the value of the variable (if percentage is intended a simple rule of three can be applied. If several reviewers assess the quality of a model, then it is recommended to either apply inter-reviewer agreement protocols or to obtain the mean value for each quality variable.

### 4.5.2. Problem 1/2/3 quality assessment: list of statements (STA)

The template consists of a tree of statements and substatements about the problem domain. The evaluator has to inspect the conceptual model so as to assess whether the substatements are included in the model or not. Each substatement has hints for the evaluator (in order to facilitate the conceptual model inspection). Figure 31 is an example correction template for the conceptual model shown in Figure 33, in order to clarify this idea (it is not structured as a tree but as a list and it does not include hints).

| Statement | The statement is... | | |
|---|---|---|---|
| How easy it was to check this statement in the model?  1=Very easy    7=Very difficult | Correctly specified in the model | Incorrectly specified in the model | Not in the model |
| 1. A car has at most four wheels  1☐ 2☒ 3☐ 4☐ 5☐ 6☐ 7☐ | ☐ | ☒ | ☐ |
| 2. A wheel can only be part of one car  1☒ 2☐ 3☐ 4☐ 5☐ 6☐ 7☐ | ☒ | ☐ | ☐ |
| 2. A car has the following states: Working, Broken  1☐ 2☐ 3☐ 4☐ 5☒ 6☐ 7☐ | ☐ | ☐ | ☒ |

Figure 31.   Example of a conceptual model correction template based on a list of statements

Then the variables are computed this way:

- *Completeness_STA* = Sum of correctly specified statements / Number of statements
- *Validity_STA* = 1 – (Sum of incorrectly specified statements / Number of statements)
- *Comprehension_STA* = Mean of the Likert scale values (only considering correct and incorrect statements)

If several reviewers assess the quality of a model, then it is recommended to either apply inter-reviewer agreement protocols or to obtain the mean value for each quality variable.

The correction template is derived from the requirements specification of the information system. Figure 32 shows a partial view of the Microsoft Excel correction template for Problem 2. This template was used in the pilot experiment; it does not include a Likert scale and it implements the measurement of validity in a different way.

Figure 32.   Partial view of the correction template for Problem 2

For ach substatement, the reviewer assesses whether the substatement has been specified in the model and assigns a value between 0 and 1 to column Compl. In case any justification for the assigned value is needed, it is recommended to write a comment in the cell.

Also, the number of validity errors encountered while assessing the substatement in the subject's model (if any) is noted down in column Valid. For each validity error, it is recomended to write a comment in the cell explaining the error.

Total amounts are displayed on top. For instance, the model by Subject 1 has an degree of semantic completeness of 87,24% and 6 validity errors.

The reviewer should note down the amount of time spent on assessing the model. A button labelled *Now* automatically timestamps the active cell; if the cells labelled *Start time* and *End time* then the cell *Total time* displays the amount of time. Another option is using an online chronometer and writing down the value in the template.

### 4.5.3. Problem 1/2/3 quality assessment: list of questions (QUE)

This instrument has not been developed yet. The following is just an example, in order to clarify this approach.

Let's suppose the following problem definition: "A car is identified by its license number. We are interested in the colour of the car. A car always has four wheels. Each wheel can be part of only one car. Each wheel is identified by a number engraved in the tyre. We are interested in the brand of each wheel. Additionally, the car can be working properly, or it can be broken down; we are interested in knowing the state of a car. When a car is broken, it is sent to a mechanic; we are interested in knowing which mechanic is fixing the car."

Let's suppose that a student has created the following conceptual model.

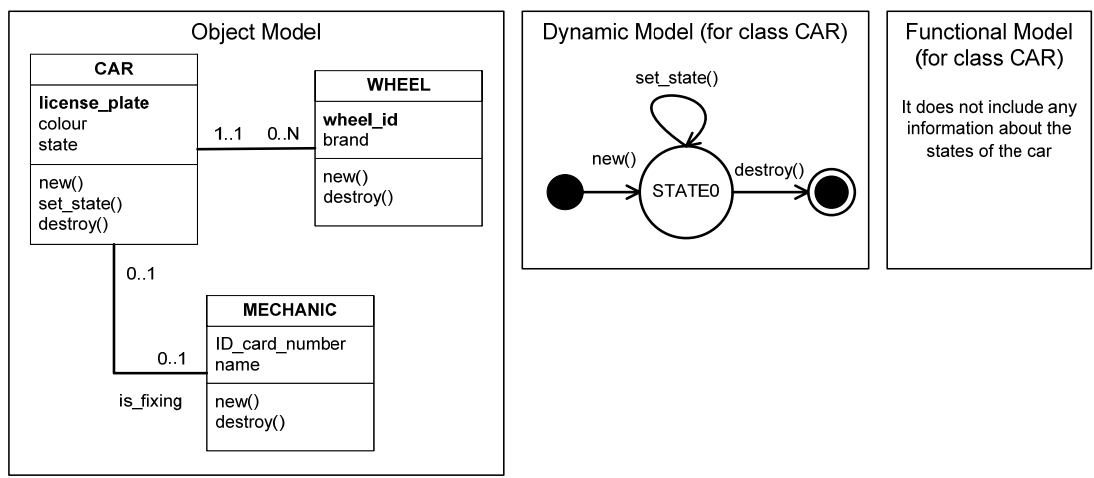

Figure 33. Example of a simple conceptual model

The reviewer is given a list of questions about the model and s/he has to answer them according to what the model indicates. Then a meta-reviewer checks whether the answer coincides with the correct value to that question.

| Reviewer | | | Meta-reviewer | | |
|---|---|---|---|---|---|
| Question | Answer | Correct value | The answer is... | | |
| How easy it was to find the answer?  1=Very easy    7=Very difficult | | | Correct | Incorrect | Not in the model |
| 1. Which is the maximum number of tyres a | N | 4 | ☐ | ☒ | ☐ |

| car can have? | | | | | |
|---|---|---|---|---|---|
| 1☐ 2☒ 3☐ 4☐ 5☐ 6☐ 7☐ | | | | | |
| 2. Of how many cars can a wheel be part of? | 1 | 1 | ☒ | ☐ | ☐ |
| 1☒ 2☐ 3☐ 4☐ 5☐ 6☐ 7☐ | | | | | |
| 3. What states can a car have? | Not in the model | Working Broken | ☐ | ☐ | ☒ |
| 1☐ 2☐ 3☐ 4☐ 5☒ 6☐ 7☐ | | | | | |

Figure 34.  Example of a conceptual model correction template based on a list of questions

Then the variables are computed this way:

- *Validity_QUE* = 1 – (Sum of incorrectly answered questions / Number of questions)
- *Completeness_QUE* = Sum of correctly answered questions / Number of questions
- *Comprehension_QUE* = Mean of the Likert scale values (considering only correct and incorrect answers, not those absent in the model)

If several reviewers assess the quality of a model, then it is recommended to either apply inter-reviewer agreement protocols or to obtain the mean value for each quality variable.

It can still happen that the reviewer interprets the model wrongly and rates the question as "easy to find the answer".

| 2. Of how many cars can a wheel be part of? | N | 1 | ☐ | ☒ | ☐ |
|---|---|---|---|---|---|
| 1☒ 2☐ 3☐ 4☐ 5☐ 6☐ 7☐ | | | | | |

Figure 35.  Example of a wrong assessment by a reviewer

This would affect completeness and validity, but more importantly, comprehension. The fact that the reviewer got the wrong answer implies that the conceptual model is not easy to understand.

# 5. EXPERIMENT OPERATION, RESULTS AND LESSONS LEARNED

This section presents the experiment log; that is, the annotations made by the researcher during the experiment operation. The results are discussed; since there were only 3 subjects that completed the course/experiment, no statistical analysis can be carried out on the collected data. We show some data (mainly raw data, but sometimes aggregated data, especially in surveys) and comment some qualitative information. Also, some lessons learned are discussed.

## 5.1. Preparation of the experiment and the course

Since the experiment is quite long (the subjects are involved for at least 2 months), we found that we had to either pay students or professional practitioners for their participation in the experiment or set the experiment within a course and have the students be the subjects. We calculated that, given the amount of variables, we needed at least 30 subjects to ensure results of statistical significance (less if we only intend to gather qualitative information). Therefore, due to budgetary constraints, the experiment was set up as part of an optional course in University of Twente.

The course was offered to 2nd and 3rd year students of the Bachelor in Business and Information Technology (BIT) and the Bachelor in Computer Science (INF), so we had 200 potential experimental subjects.

|  | INF | BIT |
|---|---|---|
| **First-year students** | N/A | 36 |
| **Second-year students** | 46 | 18 |
| **Third-year students** | 45 | 91[16] |
| **Potential enrolments** | 91 | 109 |
| **Total potential enrolments** | 200 ||

Table 7. Figures of potential enrolments in the year 2009

The news about the course were spread also by word of mouth and we ended up with some enrolments from the Bachelor in Industrial Engineering and Management (TBK).

The course was advertised by several means: a short description in the campus newspaper UT Nieuws (see Figure 37), a poster that was fixed in two or three corridors of the computer science department (see Figure 36), emails to students using course distribution lists, verbal announcement in a lecture. We also created a course site using Blackboard, a digital learning environment with which at University of Twente, in which we provided a description of the course, we made course material available as the course progressed.

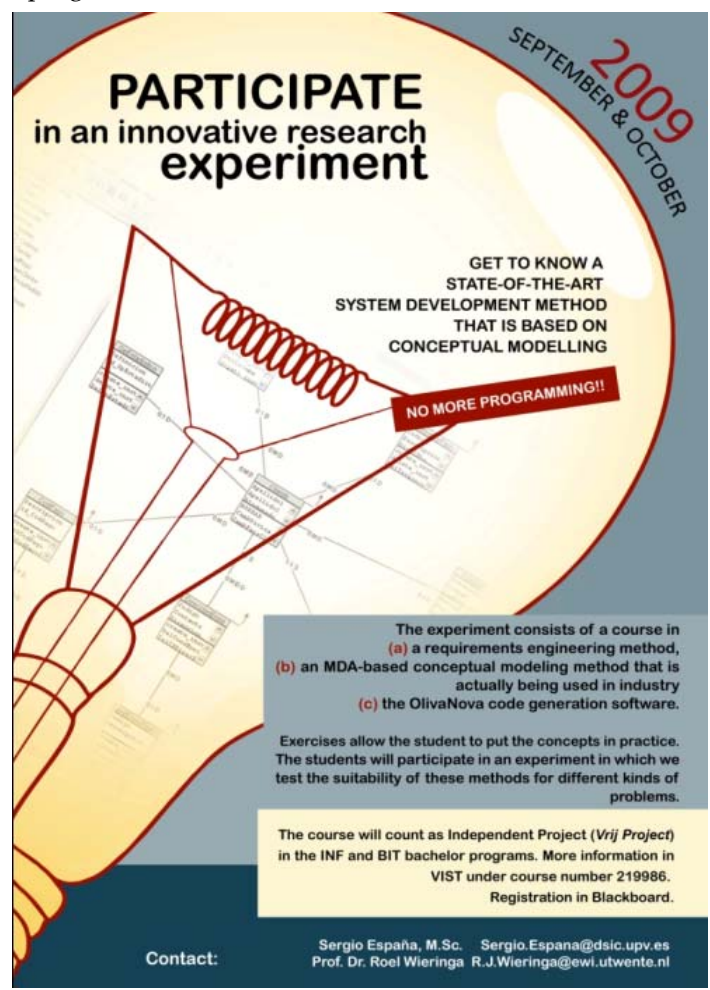

---

[16] BIT actually reported having 22 third-year students and 69 students that started studying before September 2007.

Figure 36. Course advertisement

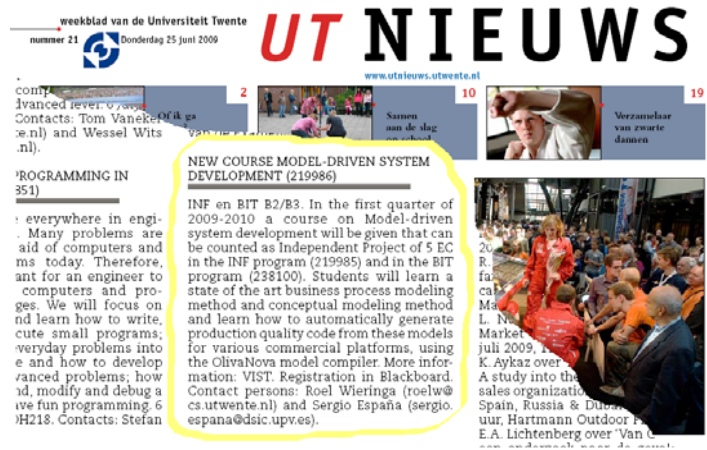

Figure 37. Composition that shows the front page and course advertisement in the newspaper

Although there seemed to be interest among students, they were reluctant to enrol the course. Main showstopper was Friday evening sessions (most students travel back to their parents' house on Fridays), so we changed this sessions to Thursday evening. Also, a late evening schedule is not convenient for students, but we could not arrange a different schedule. At the start of the course, we had 10 students enrolled. Four sessions later the number dropped to 5 students. Known reasons for dropouts are: lack of interest in the model-driven development paradigm (some students declare themselves born programmers), problems with the schedule (incompatibility with other activities; e.g. sports), and lack of interest in the information systems domain (e.g. one student preferred video game development).

We involved CARE Technologies in the experiment. A proffessional trainer was invited to train the students in the OO-Method and the *OLIVA****NOVA*** technology.

## 5.2. Kick-off

We presented the course, including an overview of its content and the agenda.

The in-take assessment took more than planned (a mean time of 30 minutes, when we had estimated 15 minutes). We realised that the survey should include questions about their industrial experience and a list of related courses that they may have attended before, in order to elicit qualitative information about their previous experience in information systems analysis and design. Some subjects have not pressed the final submit button and the survey remains uncompleted (no access to the data with a free SurveyGizmo account). It is therefore important to insist in the need of pressing the button, and to check in real-time how many responses have been collected at the end of the assessment so as to identify those subjects that have not pressed the button.

Modelling exercises take longer than expected too. Most students did not finish them; some only drew some doodles. Also, we should have given them a printed copy of the problem statement, which comes in handier than reading it in the screen.

Given the importance of the training in the experimental results, closer monitoring of students is needed during the course (e.g. to do a roll call to check whether they miss any lectures, to mark in real time that they hand in each and every deliverable –surveys, models, etc-). We recommend a paper notebook with a record of each subject (including a photo), where quick comments can be added.

**Results of the in-take survey**

The in-take assessment was carried out by 9 subjects (first nine data-rows; we only identify those that did not drop out). As shown in Section 4.1.1, the in-take survey had three groups of questions. Table 8 shows the mean values of these groups of questions for each subject.

| Subject | Knowledge of the syntax of analysis techniques | Experience in applying analysis techniques | Intention to use analysis techniques |
|---|---|---|---|
| 3 | 3,0 | 3,0 | 5,0 |
| 4 | 4,0 | 4,0 | 6,4 |
|  | 2,0 | 1,9 | 4,0 |
|  | 1,3 | 1,4 | 3,6 |
|  | 3,5 | 3,1 | 4,6 |
| 1 | 2,5 | 2,3 | 5,4 |
| 2 | 2,4 | 2,6 | 4,4 |
|  | 3,4 | 3,5 | 5,3 |
|  | 2,6 | 2,4 | 3,8 |
| Reference | 4,3 | 3,4 | 5,0 |

Table 8. Results of the in-take survey

Interestingly, Subject 1 rated his own experience low in some of the techniques but had actually a high modelling competence. In contrast, Subject 4 rated his knowledge and experience quite high in comparison to his later performance during the course. These results warn against taking user perceptions too seriously, since they are very subjective and seem to depend on factors that are variable upon subjects (e.g. self-confidence, pride, humility).

Just as a reference, the last data-row corresponds to an lecturer from University of Twente (he was not an experimental subject), with actual high knowledge of many of the syntaxes, a demonstrated industrial experience in many of the techniques, and the conviction that system analysis techniques are of great value.

## 5.3. Training on OO-Method / OLIVANOVA

### 5.3.1. Training sessions

The subjects have received 14 hours of training on the OO-Method. An expert trainer from CARE Technologies has been involved in the course and the official training material has been used(see Section 4.2.1). The subjects have been trained in the modelling techniques (Object Model, Functional Model and Dynamic Model) and in using the *OLIVA**NOVA*** technology (especially the Modeler). The training sessions were successful; the subjects were able to learn the concepts and discuss them with the trainer. They managed to use the *OLIVA**NOVA*** Modeler to carry out some exercises. In any case, there is some space for improvement, in case the experience is repeated in the future.

- A brief introduction to information systems would be convenirent before entering the *OLIVA**NOVA*** training, which is very tool-oriented. However, the content of the introduction depends on the previous knowledge of the subjects.

- Showing an automatically generated application in the first session could hook some students. We didn't show any until the fourth session.

- CARE training is mainly focused on specifying conceptual models, and not so much in conceiving them. Perhaps a lecture for teaching the students how to "reason" their models could help. All in all, full training needs more time, especially for the exercises (at the end of each session, students have not finished the exercise). Given the intensity of the training (sessions were very close to one another due to budgetary constraints) we did not deliver homework; perhaps we should have but it is doubtful that the students would have found time.

- The training could include some sort of basic requirements elicitation approach, even if it is made of a checklist of questions that the subjects should ask in order to create the conceptual model.
- At the end of the last *OLIVANOVA* training session, we briefly explained the modelling task and its instructions, including how the helpdesk-based elicitation worked. However, it would be convenient to have them place a question or two in the helpdesk to avoid later troubles.

### 5.3.2. *OLIVANOVA* modelling competence assessment

We alloted 0:30 for the knowledge test (one minute per question) and 1:45 for the (simplified) Containers modelling exercise. Some subjects are not able to finish the exercise on time. Subjects could not consult the course material during the test but could indeed do it during the exercise.

Model quality assessment was done using a template based on statements about the domain (see Figure 38). It is a cumbersome task (a mean time of 1:20, one hour and twenty minutes, per model) but models can be assessed this way (e.g. subject 2 has a degree of completeness of 92,6%). See further discussions on the correction templates in Section 5.10.

Figure 38. Fragment of the correction template for the Containers case

## 5.4. Resolution of Problem 1

For three days after the delivery of task instructions no questions were placed in the helpdesk. Then they started placing requests. As expected, the subjects asked multiple questions in one request. Subject 1, for instance, grouped them by common subject.

Some subjects used unexpected features such as starting a *discussion* within a helpdesk request (a feature of the helpdesk by which a conversation can be maintained without entering new requests). For instance, when I asked Subject 4 about the meaning of request 26, he started a conversation that we later converted into different requests (this way a many to many relationship between questions and requests is avoided). But this issues need to be further investigated if the helpdesk elicitation is to be improved for future experiments. It may involve reconsidering the conceptual model in Figure 18 (e.g. a question can actually be viewed as a dialog that actually involves communication in both directions or a dialog can be considered to contain several questions)

There are big differences in the performance of different subjects. E.g. Subject 1 places very accurate questions, others such as Subject 4 have more trouble asking proper requirements elicitation questions; Table 9 shows two questions that are inappropriate to adress a domain expert who is not an expert in the technology, as well as the answers we gave.

| |
|---|
| **Q62,** 26, Subject 4<br>What are the rights of each of the members of the Office Staff, that are not described in the problem description? |
| What do you mean by rights? |
| **Q63,** 32, Subject 4<br>By Rights I mean the visibility of attributes and services for all the users of the system. |
| I do not know what you mean by attribute or by service. I am just an electrical engineer.<br>What exactly about our business do you want to know? |

Table 9. Examples of inappropriately-formulated questions

## 5.5. OLIVA*NOVA* doubt solving session

The session had the following structure:
- Interviews
- Discuss solution to OLIVA*NOVA* knowledge test
- Discuss solution to OLIVA*NOVA* conceptual modelling exercise (containers).
- Discuss solution to Problem 1 (projects office).

### 5.5.1. Interviews

We asked a few questions to Subject 1 in private, one of the high modelling-competence subjects.
- He had previous requirements engineering experience because he has his own software development business and he has to deal with ill-defined requirements all the time.
- When he reaches a certain stage of elicitation, he starts asking questions in the form "Is it correct that...?" in order to ascertain his assumptions, and also to use their responses as an informal "contract".
- He felt that we were "hiding some information all the time". E.g. when he asked about exceptional behaviour we did not mention the worker and engineer records (actually, from the point of view of Communication Analysis, this is not exceptional behaviour but simple CRUD data management).

We later had an open interview with Subject 1, Subject 2, Subject 3 and Subject 4 in which the following topics were discussed.

**Subjects' impressions of the OLIVA*NOVA* training**
- They consider the Object Model to be easy for them, since they have previous knowledge of class diagrams.
- They find services and transactions more difficult. E.g. the script language is new for them (Subject 4); they find it a bit awkward (Subject 3 complains about the dots for parsing sentences). Subject 1 finds the collection operators impressively useful, though.
- Subject 4 asked for more material about transactions. Others would also appreciate it.

- They did not print the course material. Subject 1 prefers to use it in a digital form; he would have taken notes during the lectures but decided not to do it because otherwise he would have had trouble listening to "the trainer's soft voice". Subject 2, Subject 3 and Subject 4 tried to print it but they had trouble ("the printer got blocked"). Subject 3 suggested that the problem might be the background image of the slides.
- Some of them did not have enough time to complete the *OLIVANOVA* assessment modelling exercise (Subject 3).

**Subjects' impressions about the course, in general**

- They feel they need individual feedback. During the course they were showed reference models, but they need to individually discuss their own models with us. We agreed to make individual appointments during week 4.
- In the future, we should plan enough time to have individual meetings with the students *before* they undertake the modelling tasks.
- They would like to be allowed to use code generation services.
- They express that the most difficult part of the OO-Method for them is the Functional Model and the specification of transactions. In future experiments, we should consider extending the *OLIVANOVA* training on this part of the conceptual model and also extend the guidelines for its derivation from the requirements model.

**Subjects' impressions about conceptual modelling task Problem 1**

- Subject 3 found the setup quite realistic. He sees the fact that he had to place questions as a good surrogate for the interviews.
- It seems to be a general impression that textual specifications are not realistic.
- Subject 4 has had a course on requirements engineering where students are split into teams and given a short problem statement. Then they can prepare some questions and carry on an interview with a customer (role play). But then, unlike in our course, there was no time for follow-up elicitation interviews.
- Several students have expressed that they appreciate the quick response of the questions they placed in the servicedesk. They sense that it must be difficult for us to be always available online. Subject 3 commented that perhaps we need a tailor made tool to support elicitation.

In general, they feared that their grade depended on exercises that they are asked to do before they receive the necessary training; that is, they first do one or two exercises and then they receive requirements engineering training. We explained them that we were going to take that into account (see Section 6.5).

## 5.5.2. Doubts and comments about the OO-Method

Only Subject 1 and Subject 4 placed methodological doubts via the helpdesk. In all cases, the response that were sent was "We have taken note of your doubt. It will be solved in the next session. For the moment, make your guess and try to solve the issue with your current knowledge of *OLIVANOVA*."

> **MethQue1**   30,   Subject 4
> In the problem description are 2 project trays mentioned.
> How can I implement such trays? Will it work when I create 2 boolean-attributes for my class ProjectRecord; NewProject (boolean) & AssignedProject (boolean)?
>
> **MethQue2**   65,   Subject 4
> I'm not sure I understand the agent-visibility of Roles. So I hope it is possible to get a short explanation of the use of adding one or more roles to an agent.
>
> **MethQue3**   67,   Subject 1
> Why isn't this allowed as a creation transaction?
> Customer.create_instance(pt_p_atrid_Customer, pt_p_atrcompanyName,

```
            pt_p_atraddress, pt_p_atrVATNumber, pt_p_atrcontactPerson,
            pt_p_atrcontactTelephone)
            .
            create_instance(pt_p_agrEngineer, pt_p_agrCustomer, pt_p_atrprojectId,
            pt_p_atrprojectDescription).
```

Table 10. Methodological doubts places via the helpdesk during Problem 1

Subject 3 expressed the feeling that the state-transition diagrams of the Dynamic Model should offer a way to hide the some transitions (those that do not correspond to observable behaviour; mainly those that are not communicative events). And that internal and external services should be somehow better differentiated in the framework. It is interesting that this is in line with Communication Analysis modelling levels of emergence.

## 5.6. Allocation of subjects to groups

Figure 39 shows the procedure we followed to allocate the subjects to the treatment groups. We now comment some issues.

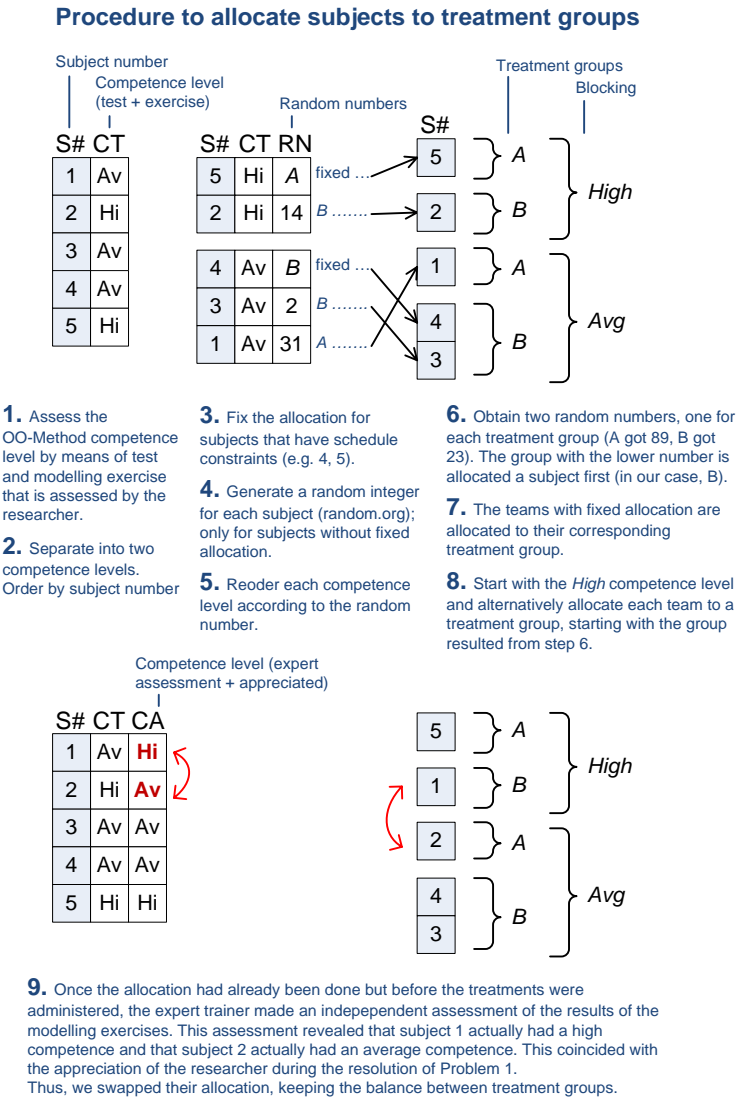

Figure 39. Allocation procedure

For random number generation we used a simple "True Random Number Generator" service available at RANDOM.ORG http://www.random.org (we executed the service 7 times, with the parameters set at Min=1 and Max=100). However, by using this service, it is possible to get duplicate numbers. In order to avoid duplicates, the "Random Sequence Generator" service is recommended.

The modelling competence assessment did not take into account how well they managed to elicit requirements; it just focused on how well they managed the tool and how well they translated a detailed object-oriented textual requirements specification into an *OLIVA**NOVA*** conceptual model. Therefore it was not a good estimator of how well they performed during the first conceptual modelling task (Problem 1). In fact, a student that was initially considered to have a high modelling competence ended up not performing too well (subject 2). And a student that was initially considered to have a medium-high modelling competence ended up being quite an expert (subject 1). In fact, his smart requirements elicitation questions led me to ask him if he had previous experience in real projects; he told me he has his own software development company.

Moreover, the industrial trainer inspected the conceptual models of the knowledge assessment and confirmed the above-mentioned appreciations. This definitely means that the correction template is not very good (or that my interpretations while applying the template where wrong). The initial random allocation of students to groups had been based in inaccurate data. Thus, in view of the new evidences, we tweaked the allocation by swapping the subjects.

## 5.7. Training on Communication Analysis

### 5.7.1. Training sessions

The subjects have received 16 hours of training on Communication Analysis. The trainer has been a researcher who is expert in the method and has applied it in industrial settings for 2 years. The training material is described in Section 4.4.1. The training sessions were successful; the subjects participated actively and improved their requirements engineering habilities. In the following, we enumerate some observations.

**Subjects' impressions of the Communication Analysis training**

- Informing the students of the degree of completeness of their models for the first assignment would be a good motivation for this part of the course.
- The subjects ask whether they could use other notations for creating business process models (other than Communicative Event Diagrams, for instance BPMN). At first, we saw no problem as far as they modelled everything that was necessary (e.g. primary actors, communicative interactions, receiver actors). But then we realised that:
    - It is too early for them to adapt other notations; they do not have enough critical eye yet.
    - Most notations usually have some underlying criteria that is somehow incompatible with the communicational view. E.g. BPMN organises activities in lanes using as criteria who carries out the activity, whereas in the Communictive Event Diagram, the important criteria is who is the primary actor (the provider of information).
  
  Therefore, we told them to stick to the Communictive Event Diagram for the logical events and use any other notation for the physical events.
- The subjects express that the way of structuring requirements in layers seems useful to them, but this kind of comments could be intended to please the teacher.
- Subject 1, who is by far the most competent student, expresses often that he appreciates the fact that we insist in the underlying criteria and systematic guidelines and not in the notation. During an exercise done in the last Communication Analysis training session, he is the only one that adopted the systematic approach for deriving the conceptual model from the requirements specification (by means of integrating class diagram views); Subject 3 and Subject 4 started drawing unrelated classes with intuition as their guiding criteria, and they left for a second iteration the check of the appropriateness of the drawn classes).

**Experimenters' impression of the subjects' attitude**

- When we sent homework, the students would not do it because it is not being graded. In any case, they did engage in the experimental tasks.
- One student (not from our course) has reported that many students just want to get a 6 in their grades and that they will do the least effort that is necessary to achieve that grade. Also, that many students share solutions of assignments in a cooperative way. However, perhaps due to the optional nature of the course and the unusual schedule, we got subjects that seemed to be very motivated by the course and participated actively in it.

### 5.7.2. Communication Analysis modelling competence assessment

The test was not appropriate for measuring the modelling competence because it only covered two of the six educational objectives [Bloom, Engelhart et al. 1956] (namely; knowledge and comprehension); also, the test had not been pre-tested and it happened to be too focused on terminology. The subjects have trouble with all the new terminology, which does not mean that they cannot apply the techniques properly if they have the method documentation available. In any case, all subjects passed the test by scoring 13 or above (over 24 questions). Also, the test allowed identifying some misconceptions that were clarified before the experimental tasks.

For future experiments, it It would also be interesting to have a small case to be modelled with Communication Analysis, as part of the modelling competence assessment (as in the OO-Method modelling competence asessment).

During the Communication Analysis training it is also possible to collect some models to monitor their performance (during the pilot experiment, we collected the conceptual model that three subjects derived from a reference requirements specification as part of an exercise, and a partially complete requirements model created by one of them)

## 5.8. Resolution of Problem 2

Following the example shown in Section 3.3, we include an example of how requests placed by the students are split into questions, how they relate to the question-answer catalogue, and how the asnwer to the request is composed of the asnwers to the questions. In the following, two statements about the domain are shown (Sta53 and Sta61), as well as three requests placed by the subjects (requests 70, 111 and 132) and the questions in which the requests have been split (Q85, Q86, Q140, Q141 and Q156). The templates differ from the ones that were used in the pilot experiment (see e.g. Figure 29), but are intended to improve the readability of the example.

| **StatementID:** Sta53 | **Title:** Publishing house requests an order. Support actor and communication channel |
|---|---|
| **QA catalog index:** | PHO 6. Publishing house orders report<br>→ 2 Contact requirements<br>→ 2.1 Actor responsibilities |
| **StatementText:** | When a publishing house is interested in a report about a specific matter or event, they phone the agency asking for information (currently it is the only channel of communication that they use). A Production Department clerk attends to the publishing houses, searches in the report filing cabinet, selects |
| | the reports that the clerk considers more interesting and suitable, and faxes a listing of the reports and their description. The publishing house requests the chosen reports over the phone, again. |

| | | | | |
|---|---|---|---|---|
| **StatementID:** | Sta61 | **Title:** | | Publishing house requests an order. Message content. Not to include listing |
| **QA catalog index:** | PHO 6. Publishing house orders report<br>→ 3 Communication requirements<br>→ 3.1 Message structure | | | |
| **StatementText:** | The list of reports that was faxed to the publishing house is actually thrown away; it is not stored. This information calls are not recorded. | | | |

| | | | | | |
|---|---|---|---|---|---|
| **RequestID:** | 70 | **SubjectID:** | 4 | **Title:** | Publishing house, interested in a report |
| **Creation date:** | 10-05-2009 18:08 | | **Completion date:** | 10-05-2009 21:14 | |
| **Request text:** | When a publishing house is interested in a report, do they always phone the agency to ask for information? When a publishing house calls to ask for information, does the Production Department makes a note of this call and of the information he/she has sent to the publishing house? | | | | |
| **Request response:** | Currently, yes. When a publishing house is interested in a report about a specific matter or event, they phone the agency asking for information. A Production Department clerk attends to the publishing houses, searches in the report filing cabinet, selects the reports that the clerk considers more interesting and suitable, and faxes a listing of the reports and their description. The publishing house requests the chosen reports over the phone, again. The list of reports that was faxed to the publishing house is actually thrown away; it is not stored. This information calls are not recorded. | | | | |
| **RequestID:** | 111 | **SubjectID:** | 3 | **Title:** | Requested reports information |
| **Creation date:** | 10-11-2009 14:34 | | **Completion date:** | 10-11-2009 16:56 | |
| **Request text:** | When a publishing house is interested in a report about a specific subject, they contact the agency by phone. In return the production department clerk sends a list of reports that match the subject. Is the production department clerk also the person that receives the call? And is the information about the request stored? With request I mean the subjects etc the publishing house is interested in. | | | | |
| **Request response:** | When a publishing house is interested in a report about a specific matter or event, they phone the agency asking for information (currently it is the only channel of communication that they use). A Production Department clerk attends to the publishing houses, searches in the report filing cabinet, selects the reports that the clerk considers more interesting and suitable, and faxes a listing of the reports and their description. The publishing house requests the chosen reports over the phone, again. The list of reports that was faxed to the publishing house is actually thrown away; it is not stored. This information calls are not recorded. | | | | |
| **RequestID:** | 132 | **SubjectID:** | 1 | **Title:** | list of reports |
| **Creation date:** | 10-11-2009 21:12 | | **Completion date:** | 10-12-2009 00:24 | |
| **Request text:** | The problem description mentions a publishing house can view a list of reports in some way. Is this referring to a list of reports the clerk mentions on the phone, or is there some other way a publishing house can get a list of available (or new) reports? | | | | |
| **Request response:** | When a publishing house is interested in a report about a specific matter or event, they phone the agency asking for information (currently it is the only channel of communication that they use). A Production Department clerk attends to the publishing houses, searches in the report filing cabinet, selects the reports that the clerk considers more interesting and suitable, and faxes a listing of the reports and their description. The publishing house requests the chosen reports over the phone, again. | | | | |

| **QuestionID:** | Q85 | **RequestID:** | 70 | **SubjectID:** | 4 | **StatementID:** | Sta53 |
|---|---|---|---|---|---|---|---|
| **QuestionText:** | When a publishing house is interested in a report, do they always phone the agency to ask for information? ||||||||

| **QuestionID:** | Q86 | **RequestID:** | 70 | **SubjectID:** | 4 | **StatementID:** | Sta61 |
|---|---|---|---|---|---|---|---|
| **QuestionText:** | When a publishing house calls to ask for information, does the Production Department makes a note of this call and of the information he/she has sent to the publishing house? ||||||||

| **QuestionID:** | Q140 | **RequestID:** | 111 | **SubjectID:** | 3 | **StatementID:** | Sta53 |
|---|---|---|---|---|---|---|---|
| **QuestionText:** | When a publishing house is interested in a report about a specific subject, they contact the agency by phone. In return the production department clerk sends a list of reports that match the subject.<br>Is the production department clerk also the person that receives the call? ||||||||

| **QuestionID:** | Q141 | **RequestID:** | 111 | **SubjectID:** | 3 | **StatementID:** | Sta61 |
|---|---|---|---|---|---|---|---|
| **QuestionText:** | And is the information about the request stored? With request I mean the subjects etc the publishing house is interested in. ||||||||

| **QuestionID:** | Q156 | **RequestID:** | 132 | **SubjectID:** | 1 | **StatementID:** | Sta53 |
|---|---|---|---|---|---|---|---|
| **QuestionText:** | The problem description mentions a publishing house can view a list of reports in some way. Is this referring to a list of reports the clerk mentions on the phone, or is there some other way a publishing house can get a list of available (or new) reports? ||||||||

## 5.9. Resolution of Problem 3

This case is considered by experts in the conceptual modelling technology as the most complex one. It should be handled with care. It is also a newly created case, which has still not been put in practice for teaching purposes.

## 5.10. Conceptual model quality assessment

The conceptual models have been evaluated by an experimenter, using the list of statements template (see Section 4.5.2). The degree of completeness has been thoroughly assessed by checking whether each substatement has been properly included in the model. However, the validity of the model has not been checked systematically; the protocol was that whenever a validity error was encountered, it was annotated in the template. This way, although an approximation to the validity of the model is obtained, it is likely that many validity errors remain unnoticed in the model after the assessment. The template has proved to be valuable to assess the model completeness, but the process to do so is very cumbersome due to the size of the models. Table 11 shows the time spent by the experimenter on evaluating the models, as well as the order in which the models have been evaluated and their degree of completeness (columns *O*, *Time* and *C*, respectively).

|  | **Subject 1** | | | **Subject 3** | | | **Subject 4** | | |
|---|---|---|---|---|---|---|---|---|---|
|  | O | Time | C | O | Time | C | O | Time | C |
| **Problem 1** | 2 | 2:02 | 73,24% | 3 | 1:53 | 48,68% | 1 | 1:20 | 57,79% |
| **Problem 2** | 2 | 1:14 | 88,45% | 1 | 2:08 | 87,24% | 3 | 0:53 | 69,30% |
| **Problem 3** | 1 | 3:00 | 72,57% | 2 | 2:52 | 95,50% | 3 | 0:57 | 65,43% |

Table 11. Time spent by the experimenter on evaluating the models (*O* order, *Time* h:mm, *C* completeness)

Our experience was that, the first model assessed within each problem implied a great effort. The reason is that the domain (i.e. the problem statement) has to be loaded in memory and certain decisions on whether a given domain statement can he modelled in one or several ways need to be made. Later model assessments benefit from this effort and take less time. However, it was also our impression that the more incomplete the model is, the shorter the assessment of its quality. This is straightforward because the moment the reviewer realises that a given part of the domain has not been discovered by the subject during requirements elicitation, the assessment of all its associated statements is quite fast (e.g. if a given class service has not been created, then none of its arguments have been created either). The data seems to support these impressions. The order $O$ seems to correlate possitively with *Time*, and completeness $C$ seems to correlate negatively with *Time* (more data points would be necessary to verify these hypotheses, though).

Neither the Likert scales (see Section 4.5.1) nor the list of questions (see Section 4.5.3) have been used, due to lack of time and resources. It could be interesting to test and compare the three evaluation approaches. An experiment could be set to do this: having several evaluators using one or several of the evaluation instruments. In such case, the procedure for distributing the models among the available reviewers needs to be defined a priori. For instance, each reviewer receives N models that they have to review with one quality assessment technique and N models that they have to review with another technique. They give back the results and they fill in two surveys, each one evaluating a technique (their perception about the quality assessment technique). This way we make quality assessment an experiment in itself. Inter-reviewer agreement protocols have to be considered.

## 5.11. Overall impression of the experience

### 5.11.1. Helpdesk-based elicitation

With regards to the helpdesk-based requirements elicitation, the overall impression is that it is a good surrogate for user interviews. The subjects liked the interaction and the results were successfull in terms of fluency and quality of the models. However, some issues could be improved and some require further investigation (they may be the object of an experiment by themselves).

#### 5.11.1.1. The need for a more detailed protocol for the helpdesk-based elicitation

For instance, during the experimental tasks, we realised that, as experimenters, we needed a more detailed protocol to guide the requirements elicitation process; there is a need for rules on how to react to students questions. We now comment on some of them.

**Granularity of statements**

Evidently, the granularity of the statements (i.e. the amount of information about the domain that they contain) and the way they are phrased (e.g. whether they contain hints about exceptional organisational behaviour or not) is expected to influence the results of the requirements elicitation process. This issue could be further investigated.

**Splitting questions**

Splitting the requests into questions is aided by the catalogue of statements, which acts as the source of domain information (the source of all requirements being elicited). In our case, we used the Communication Analysis requirements model (a textual document in Microsoft Word format) both as the index of statements and as the repository of questions (see Section 4.3.6). More guidelines are needed if more than one experimenter is involved, so as to behave homogeneously.

**Specialised business forms**

Subjects may not ask explicitly for a copy of a business form because they assume it is exactly the same as another one (e.g. report record vs exclusive record, in Problem 3) but perhaps we should give them a copy when they start asking about the message content.

**Wrong assumptions about the domain**

When a subject makes an assumption that is wrong, we should offer him the correct piece of information (see question Q243 in Table 12). Although it is a response to a non-formulated question, not including this information in our response would be misleading. This has to be dealt with care, anyway, and use as rationale whether a real domain expert would actually correct the mistake in a similar situation.

| Sta99 MAS 7. Student enrols in master. Master Secretary informative interview |
|---|
| Q224, 194, Subject 4 |
| When students show interest in a master, and they contact the master Secretary, is this recorded in the system? |
| Q243, 217, Subject 3 |
| If not, is there any other way how students can subscribe to the master besides contacting the master secretary? |
| The students that are interested in the master can look at the master website or also contact the Master Secretary, who further informs them about the master (e.g. courses, relevance of external lecturers). And she informs them of the dates and place of enrolment. No information about this interview is actually recorded. |

Table 12. Subject 3 expressing a wrong assumption

Here, Subject 3 misconception was to consider that the Master Secretary enrols the students in the master, when it is actually a Students Office clerk.

**Answers to generic questions**

When a generic question-answer has been catalogued then, if other students ask about this issue, they should get this answer (that is obvious). Also, if a student comments about this issue we should also give them this answer, as long as (a) we clarify them a doubt or a misconception and (b) we are not really giving them too much information *for free*.

| Sta87 MAS 4. PEC Deputy Head resolves master proposal. Treatments. Webpage |
|---|
| Q 223, 193, Subject 4 |
| Is the webpage that is created by a clerk (of what department?) part of the scope of this assignment? |
| And if so, what does a webpage of a master look like? |
| Q242, 217, Subject 3 |
| You said that a clerk creates a webpage for the master that is published in the TUO website. Is there already a way that students can subscribe to the master via the webpage? |
| There is a Students Office clerk who knows how to use a webpage composer. If the master edition is accepted, she creates a webpage with the information about the master and then she publishes it in the TUO website. The information also appears in the TUO newspaper. However, both the informative webpage and the newspaper are out of the scope of the development project. |

Table 13. Subject 3 expressing a wrong assumption

A concise answer to Subject 3 question would be "The website does not support enrolment", but a more natural answer would be to explain that the website is just for information purposes, so I reuse the answer to a general question-answer about the website (the corresponding question would be "Can you explain to me something more about the master website (e.g. who creates it, its purpose and functionality, etc.)")

### 5.11.1.2. Analysing helpdesk-based elicitation from the subject perspective

During Problem 1, Subject 1 had an outstanding performance in comparison to the other subjects. As discussed above, he had some industrial experience in requirements engineering. He placed 26 requests in the helpdesk, which were split into 39 questions. Subjects 3 and 4 only placed 11 questions and 7 questions, respectively. The completeness of the resulting conceptual models shows some correspondence with their elicitation performance. Subject 1 has a degree of completeness of 73,24%, whereas Subjects 3 and 4 had 48,68% and 57,79%, respectively.

|  | Subject 1 | | | Subject 3 | | | Subject 4 | | |
|---|---|---|---|---|---|---|---|---|---|
|  | R | Q | C | R | Q | C | R | Q | C |
| **Problem 1** | 26 | 39 | 73,24% | 8 | 11 | 48,68% | 6 | 7 | 57,79% |
| **Problem 2** | 26 | 31 | 88,45% | 31 | 50 | 87,24% | 47 | 57 | 69,30% |
| **Problem 3** | 22 | 52 | 72,57% | 36 | 66 | 95,50% | 50 | 80 | 65,43% |
| *Total* | 74 | 122 |  | 75 | 127 |  | 103 | 144 |  |

Table 14. Requests, questions and model completeness

After Problem 1, the three subjects were trained in Communication Analysis and this resulted in the following figures. During Problem 2, Subject 1 kept performing well (31 questions and a degree of completeness of 88,45%); he formulated accurate questions that unveiled many requirements, including exceptional organisational behaviour. Subjects 3 and 4 improved their performance; they placed 50 and 57 questions, respectively. The degrees of completeness of their models was 87,24% and 69,30%, respectively. The difference with respect to Subject 1 was the accuracy of their questions; in any case, the improvement was noticeable.

By the moment Problem 3 started, the subjects were starting to be fatigued by the amount of work the course involved (as they reported in later interviews). The performance of Subject 1 dropped due to tiredness (52 questions and a degree of completeness of 72,57%). Subject 3 performed very well (66 questions and a degree of completeness of 95,50%) and Subject 4 kept a reasonable performance (80 questions and a degree of completeness of 65,43%).

We did not arrange a schedule for the elicitation; that is, we did not fix a time frame when we would be available answering questions. Therefore, the subjects placed requests along the day, at any moment they had available. The experimenter had to be aware of new requests being placed and act as quickly as possible, especially when the subject seemed to be online. However, sometimes the subjects placed requests and it was not possible to answer them promptly. The result is a variability in the response time. The histogram in Figure shows the histogram of the response time; that is, the time elapsed between the subjects placed a request and they were sent the response.

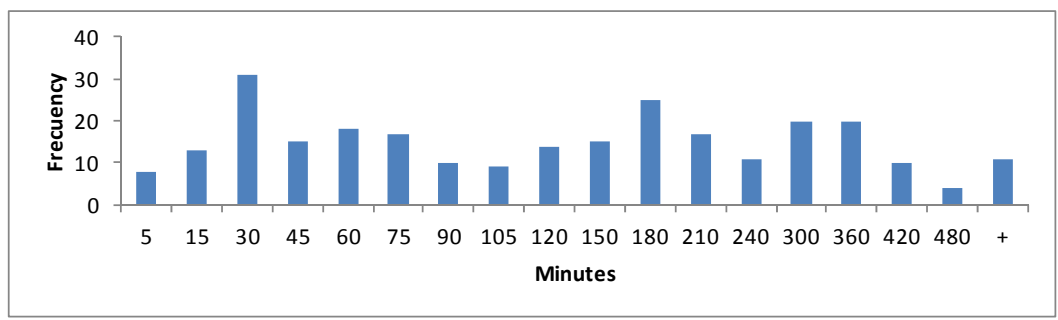

Figure 40. Histogram of the response time

The helpdesk malfunctions from time to time. It seems to have trouble with text copy-pasted from Microsoft Word. Until we figured out how to avoid that problem, some students had to re-place their questions a few times because they received tangled and unreadable answers.

### 5.11.1.3. Analysing helpdesk-based elicitation from the experimenter perspective

The experimenters prepared a question-answer catalogue in advance, before the elicitation began. However, many questions were not anticipated and, whenever a subject placed a question that was not in the catalogue, an extra effort was needed to record the question, classify it within the catalogue (i.e. place it in the Word document of the requiremetns model, in the appropriate section), decide the answer and sendit to the subject. Of course, when other subjects asked that same question later on, the experimenter only had to retrieve the answer and send it to the subject; the effort was lower. To analyse the effort involved by the help-desk based elicitation, we recorded the amount of time we spent in processing requests. We created a Microsoft Excel table (see the table that corresponds to Problem 3 in Figure 41). For each answering session (a session is a one-sit effort to asnwer questions, see columns *Se* for the session number and *Start* and *End* for the ; the session duration is calculated automatically in column *Duration*), we indicated the percentage of time devoted to the helpdesk based elicitation (sometimes an urgent task got in the middle; the estimated amount of time spent managing the helpdesk, answering questions and updating the question-answer catalogue is automatically calculated in column *Time*). The experimenter noted down the amount of requests, placed by each subject, that are attended during the session is indicated by the experimenter (see columns in the right, where R stands for requests and Q stands for questions; then the total amounts per session are automatically calculated in columns *Requests* and *Questions*).

| Se | Start | End | Duration | % dedic | Time | Requests | Questions | | Subject 1 | | Subject 2 | | Subject 3 | | Subject 4 | | Subject 5 | |
|---|---|---|---|---|---|---|---|---|---|---|---|---|---|---|---|---|---|---|
| | | | | | | | | | R | Q | R | Q | R | Q | R | Q | R | Q |
| 1 | 15-10-09 14:17 | 15-10-09 17:11 | 2:54 | 70% | 2:01 | 10 | 12 | | | | | | | | 10 | 12 | | |
| 2 | 15-10-09 17:12 | 15-10-09 18:09 | 0:56 | 100% | 0:56 | 4 | 5 | | | | | | | | 4 | 5 | | |
| 3 | 15-10-09 18:49 | 15-10-09 19:18 | 0:28 | 60% | 0:17 | 2 | 2 | | | | | | | | 2 | 2 | | |
| 4 | 15-10-09 19:19 | 15-10-09 20:46 | 1:27 | 100% | 1:27 | 11 | 12 | | | | | | | | 11 | 12 | | |
| 5 | 16-10-09 14:21 | 16-10-09 14:30 | 0:09 | 100% | 0:09 | 2 | 2 | | | | | | | | 2 | 2 | | |
| 6 | 16-10-09 15:30 | 16-10-09 16:06 | 0:35 | 100% | 0:35 | 4 | 7 | | 1 | 2 | | | | | 3 | 5 | | |
| 7 | 16-10-09 16:30 | 16-10-09 16:57 | 0:26 | 100% | 0:26 | 3 | 3 | | | | | | | | 3 | 3 | | |
| 8 | 16-10-09 21:46 | 16-10-09 23:02 | 1:16 | 80% | 1:01 | 3 | 16 | | | | | | 3 | 16 | | | | |
| 9 | 18-10-09 17:33 | 18-10-09 18:44 | 1:11 | 100% | 1:11 | 3 | 17 | | 3 | 17 | | | | | | | | |
| 10 | 18-10-09 18:44 | 18-10-09 18:55 | 0:11 | 100% | 0:11 | 2 | 14 | | 1 | 11 | | | 1 | 3 | | | | |
| 11 | 18-10-09 19:00 | 18-10-09 19:27 | 0:27 | 100% | 0:27 | 1 | 3 | | | | | | 1 | 3 | | | | |
| 12 | 18-10-09 19:27 | 18-10-09 19:40 | 0:13 | 100% | 0:13 | 2 | 4 | | 1 | 1 | | | 1 | 3 | | | | |
| 13 | 18-10-09 19:46 | 18-10-09 19:56 | 0:09 | 100% | 0:09 | 3 | 3 | | 2 | 2 | | | 1 | 1 | | | | |
| 14 | 18-10-09 20:04 | 18-10-09 20:08 | 0:04 | 100% | 0:04 | 2 | 3 | | 1 | 1 | | | 1 | 2 | | | | |
| 15 | 18-10-09 20:09 | 18-10-09 20:17 | 0:07 | 100% | 0:07 | 1 | 1 | | 1 | 1 | | | | | | | | |
| 16 | 18-10-09 20:17 | 18-10-09 20:38 | 0:20 | 100% | 0:20 | 1 | 1 | | 1 | 1 | | | | | | | | |
| 17 | 18-10-09 20:39 | 18-10-09 20:41 | 0:01 | 100% | 0:01 | 1 | 1 | | 1 | 1 | | | | | | | | |
| 18 | 18-10-09 20:42 | 18-10-09 20:58 | 0:15 | 100% | 0:15 | 1 | 3 | | 1 | 3 | | | | | | | | |
| 19 | 18-10-09 20:58 | 18-10-09 21:16 | 0:17 | 100% | 0:17 | 2 | 3 | | 2 | 3 | | | | | | | | |
| 20 | 18-10-09 21:44 | 18-10-09 22:11 | 0:26 | 100% | 0:26 | 3 | 3 | | 3 | 3 | | | | | | | | |
| 21 | 19-10-09 0:10 | 19-10-09 0:27 | 0:17 | 50% | 0:08 | 1 | 1 | | | | | | 1 | 1 | | | | |
| 22 | 19-10-09 0:28 | 19-10-09 0:38 | 0:10 | 100% | 0:10 | 3 | 4 | | | | | | 3 | 4 | | | | |
| 23 | 19-10-09 15:12 | 19-10-09 16:04 | 0:51 | 100% | 0:51 | 6 | 7 | | 2 | 2 | | | 3 | 4 | 1 | 1 | | |
| 24 | 19-10-09 16:10 | 19-10-09 16:40 | 0:29 | 100% | 0:29 | 2 | 3 | | 1 | 1 | | | | | 1 | 2 | | |
| 25 | 19-10-09 16:40 | 19-10-09 17:35 | 0:55 | 90% | 0:50 | 3 | 5 | | 1 | 3 | | | | | 2 | 2 | | |
| 26 | 19-10-09 17:37 | 19-10-09 17:54 | 0:17 | 90% | 0:15 | 1 | 1 | | | | | | | | 1 | 1 | | |
| 27 | 19-10-09 17:59 | 19-10-09 18:23 | 0:24 | 85% | 0:20 | 2 | 6 | | | | | | | | 2 | 6 | | |
| 28 | 19-10-09 19:27 | 19-10-09 19:54 | 0:27 | 100% | 0:27 | 2 | 5 | | | | | | | | 2 | 5 | | |
| 29 | 20-10-09 1:25 | 20-10-09 2:38 | 1:13 | 100% | 1:13 | 11 | 20 | | | | | | 9 | 14 | 2 | 6 | | |
| 30 | 20-10-09 13:22 | 20-10-09 13:34 | 0:12 | 100% | 0:12 | 3 | 3 | | | | | | 3 | 3 | | | | |
| 31 | 20-10-09 14:19 | 20-10-09 15:37 | 1:17 | 100% | 1:17 | 10 | 25 | | | | | | 6 | 9 | 4 | 16 | | |
| 32 | 20-10-09 16:27 | 20-10-09 16:30 | 0:02 | 100% | 0:02 | 1 | 1 | | | | | | 1 | 1 | | | | |
| 33 | 20-10-09 17:12 | 20-10-09 17:17 | 0:05 | 100% | 0:05 | 1 | 1 | | | | | | 1 | 1 | | | | |
| 34 | 20-10-09 19:52 | 20-10-09 19:54 | 0:02 | 100% | 0:02 | 1 | 1 | | | | | | 1 | 1 | | | | |
| 35 | | | 0:00 | | 0:00 | 0 | 0 | | | | | | | | | | | |
| 36 | | | 0:00 | | 0:00 | 0 | 0 | | | | | | | | | | | |
| 37 | | | 0:00 | | 0:00 | 0 | 0 | | | | | | | | | | | |
| | | Total | 18:49 | | 17:11 | 108 | 198 | | 22 | 52 | 0 | 0 | 36 | 66 | 50 | 80 | 0 | 0 |
| | | | | Time per... | | 9,5 | 5,2 | | | | | | | | | | | |

Method. doubt requests / Other requests: 2 / Total (including doubts) Requests Questions: 110 / 198

Total number of requests may not coincide with the one in helpdesk
Total number of questions should coincide with the one in the Q+A catalogue

Figure 41. Spreadsheet table that records the effort of answering helpdesk requests during Problem 3

This data allows analysing the effort graphically. Figure 42 shows the means time per question, within each answering session. We have chosen Problem 3 because the experimenter had already got the hang on the helpdesk-based elicitation process. For each session, the time spent answering questions (*Time*) is divided by the number of questions (*Questions*) effectively answered within that session. This results in a mean time that differs from session to session, depending on the difficulty of the request analysis and whether the questions that the request contains are already stored in the question-answer catalogue or not. For instance, during the first sessions, after an initial investment in order to prepare the answers to the questions formulated by the leading student, the mean time dropped because the other subjects started asking similar, foreseen, questions. Later on, during session 15, a placed a single, difficult, question that took 20 minutes to process (to focus in the answering task, analyse the request, answer it, update the catalogue, and respond via the helpdesk).

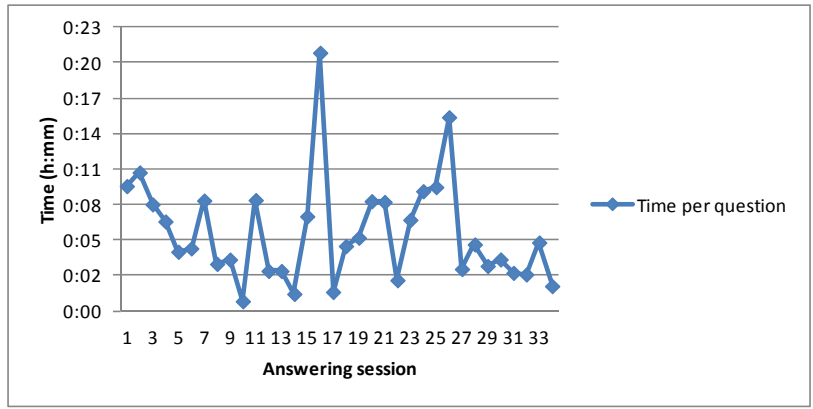

Figure 42.  Mean time per question, within each session, for Problem 3

Figure 43 shows the accumulated mean time per question. For each session, the accumulated time (the sum of *Time* from the first to the current session) is divided by the accumulated amount of answered questions (the sum of *Questions* from the first to the current session). The figure shows that, the mean time per question starts high (around 10 minutes per question) and then decreases drastically until it stabilises; the final mean time per question is 5:12 (five minutes and twelve seconds). For problems 1 and 2 the mean time is 6:49 and 5:50, respectively; this indicates that, as the experimenter had more experience with the elicitation process, he became more efficient.

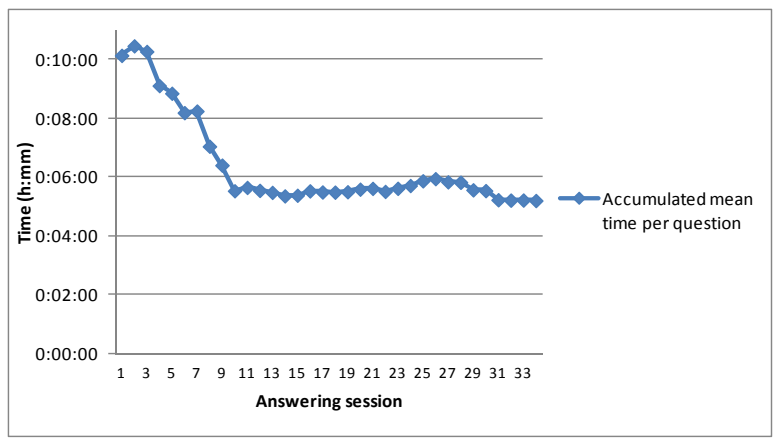

Figure 43.  Accumulated mean time per question, throughout sessions, for Problem 3

During Problem 3, the three subjects placed an amount of 198 questions (in 108 requests) via the helpdesk, which implies an average 66 questions per subject. If the group had 15 subjects (a convenient number for statistical purposes) then we can assume that they would have placed around 990 questions. Since each question took, on average, 5:12 minutes to get answered, then the experimenters would need to devote 5155 minutes (around 86 hours) attending the helpdesk. This implies an average of around 12 hours per day for a whole 7-day week. Unless more several experimenters are involved or a more automated support is offered, the helpdesk-based elicitation is not viable for bigger groups.

Some problems are related to researcher availability. We need a mechanism to receive a notification when a question is placed by a subject; otherwise the domain expert (in our case, the experimenter) needs to be looking at the helpdesk and refreshing the list of requests permanently. It could also be a good idea to agree time frames when the domain expert will be available.

A tailor-made tool to support requirements elicitation or, at least, an adaptation or the helpdesk would improve the efficiency of the experimenters. For instance, an automated support for splitting requests into questions, looking them up in the catalogue of statements and building the answer, if properly integrated with the helpdesk system, would facilitate the work of the experimenters. It should be possible to map questions to parts of the reference requirements specification easily.

### 5.11.1.4. Conceptual issues with the helpdesk-based elicitation

The conceptual model of the requirements elicitation needs more work. It could be interesting to create a conceptual model of how requirements elicitation interviews occur in practice (or adopt an existing one) and then map parts of it to the helpdesk-based requirements elicitation.

For future experiments, it could be useful to further elaborate the following aspects.

- Some factors that influence how stakeholders conceive and explain their business. E.g.:
    o expertise in the business (individual or organisational maturity)
    o expertise in technology
    o expertise in analysis (abstraction level)
    o psychological factors (such as motivation, commitment and concentration)
    o organisational factors (conflicting goals among departments)
- The emergent convenient and inconvenient behaviours (i.e. those that, respectively, facilitate or complicate the task of the analyst, such as misconceptions and miscommunications). E.g.:
    o mistakes in explanations (e.g. contradictions, inaccuracies)
    o tendency to express the business in terms of the previous software (can lead to repeating the same defects it had)
    o non-systemic approach (e.g. mixing systemic levels or granularities while describing)
    o laziness, or even boycott
    o arguments and conflicting opinions
- This would allow to empirically investigate about the abilities that the analyst should have in order to overcome inconvenient behaviours and the support that a requirements engineering technique gives to tackle the above-mentioned issues.

## 5.11.2. The involvement of the subjects in the experimental tasks

After Problems 2 and 3, we gathered data about the dedication of the subjects to the experimental task (see Table 15; RS stands for requirements engineering, CM stands for conceptual modelling)

|  | Subject 1 | | | Subject 3 | | | Subject 4 | | | Average (hh:mm) |
| --- | --- | --- | --- | --- | --- | --- | --- | --- | --- | --- |
|  | RE | CM | Both | RE | CM | Both | RE | CM | Both |  |
| **Problem 2** | - | - | 10 | 11 | 5 | 16 | 20 | 4 | 24 | 16:40 |
| **Problem 3** | 12 | 3 | 15 | 16 | 6 | 22 | 9 | 2 | 11 | 16:00 |

Table 15. Time dedication of the subjects to the experimental tasks (in hours)

The average amount of time spent by the subjects per experimental task is 16:20 (hh:mm), which is a reasonable dedication, as shown in Table 3. With respect to the distribution of time between requirements engineering and conceptual modelling, on average the subjects devoted 77% of their time to engineer requirements and 23% of their time to derive and specify conceptual model.

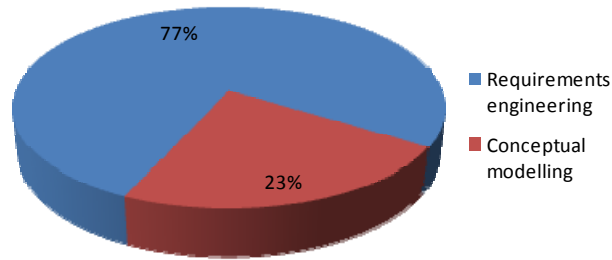

Figure 44. Distribution of time between requiremetns engineering and conceptual modelling

### 5.11.3. Quality of the resulting models

The students learnt how to create requirements models as well as conceptual models, and they were able to analyse and critisise them. The completeness of their models was not good enough to automatically generate a software application that would fully satisfy the customers. However, given their previous inexperience with the methods and the time constraints, the result is very satisfactory.

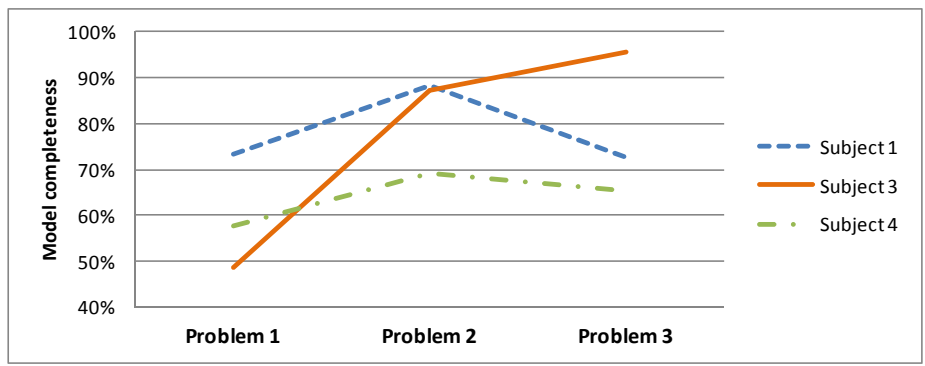

Figure 45. Evolution of model completeness throughout the experimental tasks

The evolution of the three subjects is quite different. Interestingly, we consider them to be three representatives of different phenomena that can, in pronciple, occur in an experiment such as ours. Although there is not enough data points (and no control group) to obtain significant statistical conclussions, our impressions are that the completeness of the conceptual models is related to the amount of questions that the subjects place. This relation can be seen clearly in Subject 3, who has shown a continuous increment in performance. He did not have previous experience in requirements engineering. Then, the more he learnt Communication Analysis, the better he elicited requirements and, therefore, the more complete his models were. Subject 1 was already proficient in requirements engineering. Nevertheless, as in the case of Subject 3, learning Communication Analysis helped him to improve even further his performance. His performance decrease during the last experimental task was due to fatigue; he simply quit the exercise because the reward (getting good marks) was not worth the effort (excerpt from an email: "I am aware of the deadline, but I've seen enough of Olivanova for the day"). Subject 4 did not have previous requirements engineering experience. He increased his performance after learning Communication Analysis and stabilised around 60-70% completeness. Although he made a greater effort in the last experimental task, his elicitation practice was not efficient enough. This, along with his difficulties with the conceptual modelling language, did not allow him to complete his conceptual model properly.

# 6. VALIDITY EVALUATION

In the following, we elaborate on some threats to the validity of the experiment, as discussed in [Wohlin, Runeson et al. 2000]. Also, we point out some improvements on the experimental design that would benefit the validity of the results.

## 6.1. Conclusion validity

It concerns issues that affect the ability to draw the correct conclussion about relations between the treatment and the outcome of the experiment.

A *random heterogeneity of subjects* is a risk since the variation due to individual differences can be larger than due to the treatment. To minimise this threat we have selected bachelor students, who are likely to have a similar knowledge and background. Additionally, the in-take assessment is intended to verify the homogeneity of the subjects and identify special subjects. As a trade-off, homogeneity reduces the generalisability of the conclusions (external validity).

Given the loose control that the experimenters have when the subjects are carrying out the experimental tasks (the tasks are part of their homework), *random irrelevancies* in their environment (e.g. noise, distractions, team-work) may disturb the results. However, given the length of the tasks it is the only viable experimental setting.

The *reliability of treatment implementation* depends on the similarity of the two Communication Analysis trainings. The content of the course has been established a priori; however, the students of both groups can formulate different questions or ask for further explanations on different aspects of requirements engineering. This is an unavoidable issue that, in any case, is not expected to influence significantly the results. Moreover, the Communication Analysis competence assessment (if improved, see Section 5.7.2) can be used to verify that both treatment groups end up with a similar competence of Communication Analysis after their respective training courses.

There is a threat related to the *reliability of measures*. Although the conceptual model quality framework is soundly founded in theory, our implementation of the correction template has not been tested. As part of our future work, we plan to check whether several experimenters assessing the quality of a conceptual model with the correction template based on the list of statements (see Section 4.5.2) obtain the same results (i.e. the same degree of completeness and the same number of validity errors); see Section 5.10. Moreover, having several reviewers (instead of only one) and applying an inter-reviewer agreement protocol will increase the reliability.

Also, with respect to the reliability of perception-based measures, a reliability analysis should be carried on the results of the MAM surveys using the Chronbach alpha technique. In the social psychological literature there is evidence that attitude scales, as well as skills scales, are reliable [Shaw and Wright 1967].

If we intended to draw statistical conclusions from the data gathered during the experiment operation we would face the problem of a low statistical power or that results are not significant. However, we are simply considering it as a pilot experiment that allows us to validate the experimental design and the instrumentation. In the future, if the experiment is to be operated again, a sufficient number of subjects is needed.

## 6.2. Internal validity

It is related to issues that may interfere with the treatment-outcome relationship, acting as confounding factors.

The experimental task consisting on the resolution of Problem 2 is carried out by treatment goups A and B at different moments in time (see Figures 15 and 16). This threat, referred to as *history*, is related to the fact that the circumstances may not be the same in both occasions (e.g. the students may have different levels of stress in different weeks, depending on other courses). We cannot avoid this threat but we can measure their interaction via the helpdesk, ask them if we observe anything unusual, and react according to their involvement (e.g. send remainders by email, extend deadline in case they have an exam, etc.).

There are two threats related to *maturation*. Firstly, throughout the experiment, the subjects carry out several experimental tasks (Problems 1, 2 and 3); they gradually increase their modelling competence in the OO-Method. This is expected to possitively affect the quality of their models; this way, maturation due to several applications of the OO-Method could hinder measuring the effect of applying Communication Analysis to engineer the requirements. To minimise the above-mentioned threats of history and maturation, the experiment design considers a staggered application of the treatment (i.e. Group A is trained in Communication Analysis before Problem 2, whereas Group B is trained before Problem 3; see Figure 4). This is often done in agricultural long-term experiments [McRae and Ryan 1996].

Secondly, since the experimental tasks and the overall experiment is long, the subjects can get bored or tired. This has indeed happened during the pilot experiment (Subject 1 acknowledged ending the task before solving Problem 3 completely, see Section 5.11.3). This cannot be avoided, but it can be minimised by encouraging the subjects (motivating participation) and rewarding them properly (their grades depend on the quality of their models); we can also interview them as in the pilot experiment, to find out whether they were affected by tiredness, so as to take it into account when interpreting the results.

The trade-off of our design is that the subjects in Group B have to carry out two experimental tasks consecutively. This increases the risk referred to as *resentful demoralisation*, by which a subject receiving the less desirable treatment may give up or not perform as good as it generally does. The subjects in Group B are tired after Problem 1 this may influence their performance during Problem 2. In fact, this could have been the reason behind the dropout of Subject 2 after Problem 1 (this could not be ascertained). In future replications, it is recommended to interact with the subjects of Group B before proceeding with Problem 2, in order to keep them engaged in the course-experiment.

With regards to the experiment design, it could be interesting to actually take into account the prior modelling competence of the subjects and their modelling competence in Communication Analysis. We have measured them (by means of the in-take assessment and the Communication Analysis knowledge test) but we still have not defined them as variables; also, some hypotheses related to these variables should be defined and later verified. This way, we will avoid the risk of having a hidden or *confounding factor*.

The design of the *instrumentation* affects the experiment outcome. We have learnt from previous experiments (see [España, Condori-Fernández et al. 2009; España, Condori-Fernández et al. 2010]) that paper-based surveys are error-prone. In this experiment, we have used web-based surveys to prevent blank data and transcription errors. The experience has been satisfactory in this sense.

The pilot experiment has revealed that another instrument that is capable of improvement is the Communication Analysis competence assessment. In order to effectively measure subjects competence, the following elements should be included:

- More questions in the test so as to address aspects of the method not being assessed: (i) subjects should be able to interpret a message structure; (ii) they should be able to apply Communication Analysis guidelines for model modularity.
- An elicitation exercise, so as to assess their competence in formulating questions that ask for information about an incomplete case description. The corresponding correction template/protocol is also needed.

- A modelling exercise so as to assess the requirements modelling competence of the subjects. The corresponding correction template/protocol is also needed.

There have been many student dropouts during the course-experiment; this is a threat known as *mortality*. However, we have gathered qualitative information that explains the reasons behind the dropouts (see Section 5.1); in view of the evidences, the experiment results are not compromised.

## 6.3. Construct validity

It is concerned with the relationship between theory and observation; that is, how well the treatment reflects the construct of the cause and how well the outcome reflects the construct of the effect.

We minimised the threat of *mono-operation bias* by allocating several subjects to each treatment group, and by using several problem descriptions for the experimental tasks. However, even if we are using three different cases (a projects office, a photography agency and a university master management system), the wide range of information systems is necessarily under-represented. The cases intend to be realistic and are definitively not toy examples, but their size and difficulty is still small in comparison with industrial projects.

There is a threat of *mono-method bias* with regards to the measurement of model quality, since only one way of measuring the quality of the models has been applied. Neither the Likert scales (see Section 4.5.1) nor the list of questions (see Section 4.5.3) have been used, due to lack of time and resources. This should be done in future replications. Also, it will allow comparing the three evaluation approaches, which an is interesting research. An experiment could be set to do this: having several evaluators using one or several of the evaluation instruments. In such case, the procedure for distributing the models among the available reviewers needs to be defined a priori. For instance, each reviewer receives N models that they have to review with one quality assessment technique and N models that they have to review with another technique. They give back the results and they fill in two surveys, each one evaluating a technique (their perception about the quality assessment technique). This way we make quality assessment an experiment in itself.

With regards to the *interaction of testing and treatment*, the facts that the subjects know that the experimenters assess their conceptual models and that the outcome of this assessment influences their grades is expected to influence their performance (they will presumably be motivated to produce models of high quality), but we argue that this is similar to real model-driven software development settings, where the pressure comes from the clients and from the project managers.

The *experimenters expectancies* can bias the results of an experiment. Several of the experimenters are method engineers involved in the design of Communication Analysis and indeed expect to prove the benefits of Communication Analysis. To minimise this risk, other experimenters without expectancies have been involved in this research.

We have identified three threats related to the *inadequate preoperational explication of constructs*. Firstly, with regards to the problem correction template based on the list of statements, although Lindland, Sindre and Sølvberg [1994] formalised what they meant by semantic completeness of a conceptual model, it is not clear whether all statements of a domain actually matter the same. In our template, all the (sub)statements are assigned the same weight (they contribute the same to completeness) but it is arguable whether including the registry date of an illustrated report is as important as calculating its price correctly. An open issue is how to assign weights to statements and sub-statements. A possibility is having two meta-reviewers make an initial assignment of weights (and they should also judge the template itself) and then compare their initial weights, discuss their rationale and agree the final weights.

Secondly, the short problem description (the initial description of the system given along with the task instructions at the beginning of each experimental task) is expected to influence the performance of the subjects. Depending on how much information it contains, the subjects may discover or overlook a hidden part of the system. For instance, how many business forms should the short problem description include? In the pilot experiment, we only provided some forms; perhaps one of each kind should be included. Some subjects have asked for more forms in a general way and others have asked if the organisation had a given type of forms, but then they didn't ask the same for other types of forms. In view of the questions placed via the helpdesk during Problem 2 and the resulting models, it seems that some subjects thought that the exclusive report record was the same as the regular report record. The same applies for the exclusive report delivery note and the regular report delivery note. These types of business forms are, in fact, different but the short problem description only included those of the regular report. All in all, there is a lack of theory that addresses how this issue affects the requirements elicitation and the conceptual modelling tasks.

Thirdly, the Problems 1, 2 and 3 are complex and big in comparison to other experimental tasks explained in the literature. Some subjects expressed that they did not have time to finish the assignment or that they were not willing to spend more time on them. For instance, Subject 4 expressed that he would have needed from 8 to 12 hours more to completely solve Problem 2 (half of this hours for requirements engineering and half for conceptual modelling). Subject 3 felt confident about his conceptual model for Problem 2 but reported that his requirements model was incomplete because he felt it was not worth spending more time on it (it was not being graded). However, a complex problem is needed in order to demand a strong effort from the subject and advanced features from the requirements engineering method. Otherwise the subjects would perform similarly well with any requirements engineering method, just applying common sense and standard elicitation practices (e.g. placing simplistic questions). Our impression is that there is a lack of theory about the effect of the complexity of the problems in the observability of the benefits of requirements engineering method; that is, the interaction of problem complexity and treatment. Note that complexity does not only refer to size (e.g. the amount of business objects that an information system needs to manage or the amount of business processes that are included in problem scope) but also on other factors (e.g. the fact that there exist business process specialisations and exceptional behaviours in the company).

## 6.4. External validity

It is related to the extent to which the conclussions can be generalised outside the scope of the experiment.

We are aware that the experiment would benefit by using real practitioners as experimental subjects. However, this may not be possible given the long involvement that it is required. In any case, Runeson [2003] suggested that the results obtained by using students as experimental subjects can be, to a great extent, generalised to industry practitioners. In any case, the pilot experiment did not allow to verify the hypotheses and the experiment would need to be operated again experiments with a larger number of subjects.

The complexity of the problems used for the experimental tasks is not fully comparable to real industrial problems. Nonetheless, they were thoughtfully selected because they balance complexity and feasibility (they can be tackled by students) with a limited availability.

## 6.5. Ethical concerns

From the point of view of teaching, the group to which students are allocated is expected to influence their performance and, therefore, their grades are influenced to. This has to be checked and taken care of, in case there is a significant difference in grades between students of both groups.

There exist ways to correct the difference between groups so as to grade the students fairly. We present a mechanism to increase the grades of the students in the underprivileged group so as to balance the averages of both groups.

$T$ is the set of treatments (or treatment groups); in our case, $T = \{A, B\}$.

$S$ is the set of all students, $S_A$ is the set of students in treatment group A, $S_B$ is the set of students in treatment group B.

$P$ is the set of problems; in our case, $P = \{1,2,3\}$ since there are three problems as experimental tasks.

$G_{s,p}$ is the grade of student $s$ in problem $p$ (e.g. the completeness of their model).

$$AvgG_s = \frac{\sum_{p \in P} G_{s,p}}{|P|}$$

$$GroupAvgG_t = \frac{\sum_{s \in S_t} AvgG_s}{|S_t|}$$

$$Ratio_{AB} = \frac{GroupAvgG_A}{GroupAvgG_B}$$

$$CorrectedG_s = \begin{cases} G_s & \text{if } (s \in S_A \wedge Ratio_{AB} \geq 1) \vee (s \in S_B \wedge Ratio_{AB} \leq 1) \\ G_s \times Ratio_{AB} & \text{otherwise} \end{cases}$$

Table 16 shows an example in which 10 students have been split into two groups (A and B). The average of the groups differ; that is, $AvgG_A = 4{,}93$ whereas $AvgG_B = 5{,}67$. This means that the performance of the students in group A is 0,87 times the performance of the students in group B. In case this difference is considered to be a result of the influence of the treatment, then the grades of the students in group A needs to be increased. After applying the correction factor, both groups have an average of 5,67.

| Group | Subject | Grades | | | Student average | Group average | Corrected student average |
|---|---|---|---|---|---|---|---|
| | | Problem 1 | Problem 2 | Problem 3 | | | |
| A | 1 | 7 | 8 | 7 | 7,33 | 4,93 | 8,42 |
| | 2 | 5 | 6 | 5 | 5,33 | | 6,13 |
| | 3 | 5 | 6 | 5 | 5,33 | | 6,13 |
| | 4 | 4 | 5 | 4 | 4,33 | | 4,98 |
| | 5 | 2 | 3 | 2 | 2,33 | | 2,68 |
| B | 6 | 7 | 9 | 8 | 8,00 | 5,67 | 8,00 |
| | 7 | 5 | 8 | 6 | 6,33 | | 6,33 |
| | 8 | 5 | 7 | 6 | 6,00 | | 6,00 |
| | 9 | 4 | 6 | 5 | 5,00 | | 5,00 |
| | 10 | 2 | 4 | 3 | 3,00 | | 3,00 |
| Overall average | | 4,60 | 6,20 | 5,10 | 5,30 | 0,87 | 5,67 |
| | | | | | | Ratio between averages | |

Table 16. Example of weighting students averages to correct inequality

Since the students feared that, during the first experimental tasks, they were being graded on something they had not yet been properly trained for (see Section 5.5.1), we considered the possibility of weighting the grades in a way that later exercises have more impact on the final grade that earlier ones. However, this was not necessary in the end.

Lastly, the students are aware of the fact that they are experimental subjects. It is important that they do not feel as guinea pigs, but as students. Giving them some feedback on the experiment design, the operation and the expectations of the researchers can help improving their overall perception of the experience. In fact, the three students that ended the course expressed interest in how the experiment had been designed. If experimental data and examples are shown in the explanations, their anonimity must be granted.

# 7. CONCLUSIONS

We have designed an experiment that has the capability of assessing the impact that the application of Communication Analysis during requirements engineering has on the quality of OO-Method conceptual models (see the research goals in Section 1, the design of the experiment in Section 2, its procedure in Section 3, the instrumentation in Section 4 and the validity evaluation in Section 6). The experiment has been operated in University of Twente (The Netherlands) in 2009 (see the report in Section 5). Due to the small amount of subjects, no statistical conclusions can be drawn, but the pilot experiment has proved the feasibility of the experimental design. We have collected and reported much information regarding the operation of the experiment.

We can conclude that the experiment is promising and it has the potential, not only to verify the hypothesys, but also to enlighten about many mechanisms that underlie requirements engineering and conceptual modelling.

In the future, we intend to repeat the operation of this experiment whenever the resources are available. We are of course open to external collaboration and we will provide support to replications. However, some improvements on the design and the instrumentation are recommended (see Section 5).

We have realised, however, that the experiment design is very ambitious in the sense that it covers several tasks carried out by the analysts (properly speaking, by the experimental subjects acting as surrogates for analysts); namely, requirements elicitation, requirements modelling, conceptual model derivation (the later actually includes deriving an initial conceptual model by applying transformation rules to the requirements model and later completing it by adding new elements to it).

There problem with this is that we may be able to verify the hypothesis that applying Communication Analysis to engineer requirements has a possitive effects on the quality of the OO-Method conceptual models, but we may be unable to ascertain which part of Communication Analysis (e.g. the elicitation guidelines, the modelling language, the conceptual model derivation rules) are responsible for this effect. From the qualitative point of view we can indeed draw some conclussions but the experiment does not allow this kind of analysis rom the quantitative (i.e. statistical) point of view.

This led us to design another experiment with a narrower scope, restricting itself to conceptual model derivation. It was operated in Universitat Politècnica de València (Spain) in 2010 and replicated again in 2011, but the data is still being analysed.

However, we still believe that the wider experiment described in this technical report is worth the effort, due to the rich analyses that it will allow.